\documentclass{aa}  

\usepackage{graphicx}
\usepackage{txfonts}
\usepackage[dvipsnames]{xcolor}
\usepackage[colorlinks=true,citecolor=blue]{hyperref}

\def\flamingo{\texttt{FLAMINGO}}
\def\ihMpc{h ~ \mathrm{Mpc}^{-1}}
\def\hMpc{h^{-1} ~ \mathrm{Mpc}}
\def\baccoemu{\texttt{baccoemu}}
\def\multinest{\texttt{MULTINEST}}

\begin{document}

   \title{A 1\% accurate method to include baryonic effects in galaxy-galaxy lensing models}

   \author{M. Zennaro
          \inst{1}
          ,
          G. Aricò
          \inst{2}
          ,
          C. García-García
          \inst{1}
          ,
          R. E. Angulo
          \inst{3,4}
          ,
          L. Ondaro-Mallea
          \inst{3,5}
          ,
          S. Contreras
          \inst{3}
          ,
          A. Nicola
          \inst{6}
          , 
          M. Schaller
          \inst{7,8}
          ,
          J. Schaye
          \inst{7}
          }

   \institute{
        Department of Physics (Astrophysics), University of Oxford, Denys Wilkinson Building, Keble Road, Oxford, OX1 3RH, UK\\       \email{matteo.zennaro@physics.oxford.ac.uk}    
        \and
        Institut für Astrophysik (DAP), Universität Zürich, Winterthurerstrasse 190, 8057 Zürich, Switzerland
        \and
        Donostia International Physics Center, Manuel Lardizabal Ibilbidea, 4, 20018 Donostia, Gipuzkoa, Spain
        \and
        IKERBASQUE, Basque Foundation for Science, 48013, Bilbao, Spain
        \and
        Department of Theoretical Physics, University of the Basque Country UPV/EHU, Bilbao, E-48080, Spain
        \and
        Argelander Institut für Astronomie, Universität Bonn, Auf dem Hügel 71, 53121 Bonn, Germany
        \and
        Leiden Observatory, Leiden University, PO Box 9513, 2300 RA Leiden, the Netherlands
        \and
        Lorentz Institute for Theoretical Physics, Leiden University, PO Box 9506, 2300 RA Leiden, the Netherlands
        }

   \date{Received September 15, 1996; accepted March 16, 1997}

   \titlerunning{Baryonic effects in galaxy-galaxy lensing models}
   \authorrunning{M. Zennaro et al.}

 
  \abstract
   {The clustering of galaxies and galaxy-galaxy lensing are two of the main observational probes in Stage-IV large-scale structure surveys, such as Euclid and LSST. Unfortunately, the complicated relationship between galaxies and matter greatly limits the exploitation of this data. Sophisticated theoretical galaxy bias models -- such as the hybrid Lagrangian bias expansion -- allow describing galaxy clustering down to scales as small as $k = 0.7 ~ \ihMpc$. However, the galaxy-matter cross-power spectra are already affected by baryons on these scales, directly impacting the modelling of galaxy-galaxy lensing.}
   {In this work, we propose a way to extend state-of-the-art models of the galaxy-matter cross-power spectrum $P_{\rm gm}(k)$ (currently only accounting for dark matter) by including a baryonic correction term inferred from the matter component (the suppression $S_{\rm mm}(k) = P_{\rm mm, full \, physics} / P_{\rm mm, gravity \, only} $), so that $P_{\rm gm, full \, physics} (k) = \sqrt{S_{\rm mm}} P_{\rm gm, gravity \, only}$.}
   {We use the \flamingo{} hydrodynamical simulations to measure the effect of baryons on the galaxy-matter cross-power spectrum and to assess the performance of our model.
   Specifically, we perform a Bayesian analysis of synthetic data, implementing a model based on BACCO's hybrid Lagrangian bias expansion (for the nonlinear galaxy bias) and Baryon Correction Model (for the baryon suppression of the matter power spectrum).
   }
   {Ignoring the effect of baryons on the galaxy-matter cross-power spectrum leads to a biased inference of the galaxy bias parameters, while ignoring baryons in both the galaxy-matter and matter-matter power spectra leads to a biased inference of both the galaxy bias and cosmological parameters. In contrast, our method is 1\% accurate compared to all physics variations in  \flamingo{} and on all scales described by hybrid perturbative models ($k < 0.7 ~ \ihMpc$). Moreover, our model leads to inferred bias and cosmological parameters compatible within 1$\sigma$ with their reference values. We anticipate that our method will be a promising candidate for analysing forthcoming Stage-IV survey data.}
   {}


   \maketitle
%

\section{Introduction}
Current and upcoming galaxy surveys promise to deliver unprecedentedly precise data down to extremely small scales. Developing models to describe the clustering of galaxies and matter to the smallest scales possible is essential for harnessing these data and maximising the amount of information that can be extracted.

One of the challenges in doing so is to deal with unmodeled physics and systematics. Among them, the effect of baryons on the matter distribution at small scales is a rather large uncertainty since it can induce significant modifications of the matter power spectrum (up to $\sim 20$ or $30\%$) and, at the same time, the physics governing it is not fully understood.

Hydrodynamical simulations offer a fundamental tool for shedding light on these processes and their effect on cosmological observables. However, such simulations are prohibitively expensive for performing Bayesian analyses. Moreover, a single or a few realisations cannot provide definitive results since different sub-grid models and values of galaxy formation parameters lead to dramatically different results \citep[e.g. ][]{VanDaalenEtal2011, SemboloniEtal2011, HellwingEtal2016, MummeryEtal2017, SpringelEtal2018, ChisariEtal2018, VanDaalenEtal2020}.

Many tools are available for the matter power spectrum to include baryonic effects efficiently. These range from halo model approaches \citep[e.g.][]{SemboloniEtal2011,SemboloniEtal2013,MohammedEtal2014,Fedeli2014,MeadEtal2015,DebackereEtal2020,MeadEtal2021}, approaches based on principal component analyses \citep[e.g.][]{EiflerEtal2015,HuangEtal2019}, corrections applied through machine learning tools \citep[e.g.][]{TrosterEtal2019,Villaescusa-NavarroEtal2020}, or applying Effective Field Theory \citep[][]{BragancaEtal2021}. Only recently, directly interpolating between hydrodynamical simulation outputs has become possible using emulators \citep{SchallerEtal2024b}. One of the most promising approaches is based on modifying dark-matter-only simulations to mimic the presence of baryons based on physically motivated arguments -- a class of models commonly referred to as \textit{Baryon Correction Models} (BCM) or \textit{baryonification} \citep{SchneiderTeyssier2015,SchneiderEtal2019,AricoEtal2020}. This technique, thanks to its computational efficiency, allows for the creation of large training sets to be used to build emulators of baryonic suppression as a function of cosmological parameters and a few (typically 7) physically motivated baryonic parameters \citep{AricoEtal2021b,GiriSchneider2021}. With these tools, baryonic suppressions of many different hydrodynamical simulations can be reproduced with a percentage accuracy to scales as small as $k=5~\ihMpc$.

 However, baryonic physics processes can affect not only the distribution of matter but also the distribution of biased tracers, such as galaxies. \cite{VanDaalenEtal2014} found that, in terms of two-point correlation functions, galaxy clustering on scales larger than the virial radius of haloes is not affected by baryonic effects; this means that any galaxy clustering model, provided that it is either insensitive to or corrected for modifications of the subhalo masses induced by baryons, is expected to be able to reproduce the distribution of galaxies in the presence of baryons. On smaller scales, on the other hand, these authors find that baryons do affect the galaxy distribution, since they induce modifications in the orbits of satellite subhaloes (and galaxies). Nonetheless, the effect of baryons on galaxy clustering and the galaxy-matter cross-correlation is still a commonly overlooked aspect in current cosmological analyses (such as the combined analysis of galaxy clustering, galaxy-galaxy lensing and shear, often called 3x2point). 

For the modelling of galaxy clustering, based on the results of \cite{VanDaalenEtal2014}, we can expect the effect of baryons to be under control -- specifically when using sophisticated models of galaxy bias. In fact, they are designed to absorb the complicated way galaxies track the underlying matter distribution irrespective of the details of the matter distribution itself \citep{DesjacquesJeongSchmidt2016}. Specifically, these models typically include a term accounting for non-local responses of the galaxy distribution to the matter field, and, thanks to this contribution, can accurately fit the clustering of galaxy samples created to reproduce those of hydrodynamical simulations \citep[in this case, see for example,][]{ZennaroEtal2022}. All state-of-the-art galaxy bias models, irrespective of their flavour, are designed to capture the response of galaxies to the underlying dark matter (however complicated that might be) and include higher-order terms (such as the Laplacian of the density field) that capture nonlocal effects such as those induced by baryons on the inner shape of haloes hosting galaxies. In this sense, selection effects play a more important role in galaxy clustering than baryonic effects. 

For galaxy-galaxy lensing models, however, baryons can still constitute a problem: since galaxies are not guaranteed to trace matter in the same way in the presence and absence of baryons, even if the bias parameters can absorb baryonic effects, they might assume incompatible values when applied to the galaxy auto power spectrum (or galaxy clustering) or galaxy-matter cross-spectrum (or galaxy-galaxy lensing) -- thus weakening or invalidating joint analyses of these observables \citep[see, for example, the discussion about optimistic and pessimistic forecasts for galaxy-matter cross-power spectra in the presence of baryons in][]{LesgourguesEtal2024}. Moreover, correctly accounting for the effect of baryons on the matter distribution has already been proven fundamental for understanding the origin of the so-called "lensing is low" problem \citep{ContrerasEtal2023, Chaves-MonteroEtal2023}.

This is typically not a problem if the galaxy-galaxy clustering and lensing analysis is performed on relatively large scales -- which is often the case for at least one other reason: linear bias models break down on scales of $k > 0.1 ~ \ihMpc$, and perturbative nonlinear bias models allow for including scales down to $k \sim 0.2 ~ \ihMpc$ \citep{BaumannEtal2012,BaldaufSchaanZaldarriaga2016,VlahCastorinaWhite2016,IvanovSimonovicZaldarriaga2020b,DamicoEtal2020,ColasEtal2020,NishimichiEtal2020,ChenVlahWhite2020}. For example, \cite{Andrade-OliveiraEtal2021}, analysing the DES Y1 data, limit their galaxy clustering models to $k_{\rm max} = 0.125 ~ \ihMpc$ to ensure that the linear bias description is accurate enough. Also \cite{GiriChaitanya-Tadepalli2023}, in their 3x2 point analysis of DES-Y3 data, include scales $k < 0.1 ~ \mathrm{Mpc}^{-1}$, to additionally ensure that they can avoid baryonic contamination in the galaxy matter cross-spectrum. For example, \cite{VanDaalenEtal2011} showed that baryonic effects can arise on scales as large as $0.3 ~ \ihMpc$, so it is no surprise that these analyses steer clear of such small scales. 

However, hybrid bias models have recently been introduced as a powerful tool to model biased tracers down to scales as small as $k = 0.6$ or $0.7 ~ \ihMpc$. The idea of hybrid models \citep{ModiChenWhite2020} is to combine a Lagrangian perturbative approach to expand the galaxy-matter connection, including the non-linear evolution of these fields as directly measured in simulations \citep{KokronEtal2021,ZennaroEtal2022}. These methods require building an emulator to effectively interpolate between the outputs of simulations needed to take advantage of the fully nonlinear evolution -- emulators such as the ones presented in \cite{KokronEtal2021}, \cite{ZennaroEtal2023}, or \cite{DeRoseEtal2023}, and that already are being used to analyse current data showing promising results \citep[an example being][]{HadzhiyskaEtal2021}. These models are arguably our best candidates to go to small scales and, as shown in the previous references, can handle the galaxy auto power spectrum even in the presence of baryons. However, including smaller scales means that we now have to worry about the effects of baryons on the galaxy-matter cross-power spectrum. As a matter of fact, the galaxy-matter cross-correlation might be affected by baryons non-trivially. 

In this paper, we focus on modelling the galaxy-matter cross-power spectrum in the presence of baryons (given applications for galaxy-galaxy lensing in 3x2point analyses). Specifically, we propose a simple way of including baryonic effects based on the baryonic suppression inferred from the matter auto power spectrum. We discuss the limits of applicability of this model and investigate potential biases arising from the neglect of baryons. In principle, our proposed way of including baryons can be adapted to any existing model of the suppression of the matter power spectrum, but in this work, we will focus on the BCM, given its flexibility in reproducing many diverse hydrodynamical simulations.

The paper is organised as follows: in Sec. \ref{sec:model}, we present our model for accounting for baryons in the galaxy-matter cross-power spectrum. In Sec. \ref{sec:sims}, we present the simulations we employ to validate it, including how halo and galaxy samples are selected. In Sec. \ref{sec:validation-with-sims}, we validate the model with simulations and assess its performance with haloes (\S \ref{sec:halomassbins}) and galaxies (\S \ref{sec:Ratios}). In Sec. \ref{sec:bayesian}, we discuss the details of how we perform our Bayesian analysis and investigate the effect of different assumptions for the baryonic suppression on the posterior in fixed (\S \ref{sec:fixedcosmo}) and varying (\S \ref{sec:freecosmo}, \S \ref{sec:projeffs}) cosmologies.  

\section{Modelling baryons in the galaxy-matter cross-power spectrum}\label{sec:model}
In this section, we will introduce the model we are proposing for the galaxy-matter cross-power spectrum. We will employ a hybrid model to describe the galaxy bias part and the BCM to introduce the effects of baryons. We expect the general idea to hold even when replacing the hybrid model with any other state-of-the-art galaxy bias model and the BCM with other models of the baryonic suppression of the matter power spectrum.
\subsection{Non-linear galaxy bias}
We model galaxy clustering employing the hybrid Lagrangian expansion model, in its flavour presented in \cite{ZennaroEtal2022} and based on \cite{ModiChenWhite2020} \citep[but also see][for other implementations and applications]{KokronEtal2021,HadzhiyskaEtal2021,DeRoseEtal2023,NicolaEtal2024,PellejeroEtal2023,Pellejero-IbanezEtal2024}. In this context, the galaxy overdensity in Eulerian coordinates $\boldsymbol{x}$ is described as a weighted version of the matter overdensity in Lagrangian coordinates $\boldsymbol{q}$, advected to Eulerian space with displacements $\boldsymbol{\psi}(\boldsymbol{q})$ measured in $N$-body simulations,
\begin{equation}
    1+\delta_{\rm g}(\boldsymbol{x}) = \int \mathrm{d}^3 \boldsymbol{q} ~ w(\boldsymbol{q}) ~ \delta_{\rm D}(\boldsymbol{x} - \boldsymbol{q} - \boldsymbol{\psi});
\end{equation}
here, the weights come from the expansion
\begin{equation}
    w(\boldsymbol{q}) = 1 + b_{1} \delta(\boldsymbol{q}) + b_{2} \delta^2(\boldsymbol{q}) + b_{s^2} s^2(\boldsymbol{q}) + b_{\nabla^2\delta} \nabla^2\delta(\boldsymbol{q}),
\end{equation}
with $\delta$ being the linear matter overdensity, $s^2$ the contraction of the traceless tidal tensor, and $\nabla^2\delta$ the Laplacian of the matter density field; here, $b_1, b_2, b_{s^2},$ and $b_{\nabla^2\delta}$ are free parameters -- controlling the biasing of galaxies with respect to the underlying dark matter. In this implementation, all of these Lagrangian fields are smoothed on a scale of $k_{\rm s} = 0.75 ~ \ihMpc$ to avoid exclusion effects (i.e., the fact that the Lagrangian regions corresponding to the final collapsed haloes cannot overlap); this scale, therefore, sets the limit of applicability of the model.

This expansion results in 15 cross terms $P_{i,j}$ (with $i$ and $j$ corresponding to the fields $1, \delta, \delta^2, s^2$, and $\nabla^2\delta$) to be combined, weighted by their corresponding bias parameters, to describe the clustering of biased tracers.  
The galaxy auto power spectrum is therefore described as

\begin{equation}
    P_{\rm gg}(k) = \sum_{i,j \in \{1, ~ \delta, ~ \delta^2, ~ s^2, ~ \nabla^2\delta\}} b_{i}b_{j}P_{i,j}(k) + P_{\rm stochastic}(k),
    \label{eq:Pgg}
\end{equation}
where $P_{\rm stochastic}(k)$ is, in the simplest case, a free-amplitude Poisson contribution $P_{\rm stochastic} = A_{\rm sn} / n_{\rm g}$, with $A_{\rm sn}$ being a free parameter and $n_{\rm g}$ being the number density of tracers. This is the simplest approximation for accounting for potential non-Poissonian noise (expected from exclusion effects), with $A_{\rm sn}=1$ corresponding to perfectly Poissonian stochastic noise. This can be further expanded to higher orders -- something that has proven necessary for analysing spectroscopic samples in redshift space \citep[see, for example,][]{PellejeroEtal2023,Pellejero-IbanezEtal2024}, but that we checked is not required for the samples used in this work in real space -- provided that we exclude scales where the power spectrum signal falls below 1.5 times the shot noise contribution \citep[see][for an assessment of the quality of the fits with free amplitude Poisson noise when excluding scales where the signal falls below 1.5 times the shot noise]{ZennaroEtal2022}.

The galaxy matter cross-power spectrum is therefore described as

\begin{equation}
    P_{\rm gm}(k) = \sum_{i \in \{1, ~ \delta, ~ \delta^2, ~ s^2, ~ \nabla^2\delta\}} b_i P_{1,i}(k),
    \label{eq:Pgm}
\end{equation}
where we are not including any stochastic noise.

Note that we have abused the notation for a few of the bias parameters, namely $b_{i=1} = 1$, $b_{i=\delta} = b_1$, and $b_{i=\delta^2} = b_2$.

\subsection{Baryon suppression}
In this work, we propose a simple way of modelling the effects of baryons on the galaxy-matter cross-power spectrum. In the simplest possible case, the bias relation describing the connection between galaxies and matter remains the same as in the baryonless scenario ($\delta_{\rm g} = \delta_{\rm g}(\delta_{\rm m, dmo})$), but traces a different matter distribution when baryons are present. We will then assume that $\left\langle\delta_{\rm g} \delta_{\rm m}\right\rangle_{\rm hydro} \approx \left\langle \delta_{\rm g}(\delta_{\rm m, dmo}) ~ \delta_{\rm m, hydro} \right\rangle$. We, therefore, need to model $\delta_{\rm m, hydro}$.

Many models already exist to describe the baryon suppression on the \textit{matter} power spectrum, 
\begin{equation}
S_{\rm mm}(k) = P_{\rm mm, hydro}(k) / P_{\rm mm, dmo}(k), 
\end{equation}
both based on phenomenology and physical considerations. We propose to use $\sqrt{S_{\rm mm}}$ to include baryon effects on the galaxy matter cross-power spectrum. This means that we can write the theoretical prediction for the galaxy matter cross-power spectrum as 

\begin{equation}
    P_{\rm gm, hydro} = \sqrt{S_{\rm mm}(k)} P_{\rm gm, dmo},
\end{equation}
or, equivalently, that the baryon suppression on the cross-power spectrum $S_{\rm gm} = P_{\rm gm, hydro} / P_{\rm gm, dmo}$ is well approximated by

\begin{equation}
    S_{\rm gm}(k) = \sqrt{S_{\rm mm}(k)}.
\end{equation}
This is valid only on the scales where the matter and galaxy fields can be regarded as separable, and therefore, their cross-correlation satisfies:

\begin{equation}
    \langle \delta_{\rm m} \delta_{\rm g} \rangle = \sqrt{\langle \delta_{\rm m} \delta_{\rm m} \rangle ~ \langle \delta_{\rm g} \delta_{\rm g} \rangle}.
\end{equation}

We expect that, as we enter the inner regions of dark matter haloes and galaxies begin to be affected by baryonic feedback in ways that are not entirely correlated with the underlying dark matter, this approximation will inevitably break down \citep{VanDaalenEtal2014}. However, we devote the remainder of this work to show that this simple approximation suffices on the scales described by even the most sophisticated galaxy bias models and leads to unbiased posteriors.

In this work, we model $S_{\rm mm}(k)$ using the emulator available in the \baccoemu{} suite \citep{AricoEtal2021b}. This is based on the \textit{baryonification} model (or \textit{Baryon Correction Model}, BCM), first introduced in \cite{SchneiderTeyssier2015}, but in this case, following the implementation presented in \cite{AricoEtal2020}. In the BCM, particles inside dark matter haloes are displaced according to the difference between the DMO halo profile and physically motivated profiles describing a central galaxy, the distribution of gas in the halo, and the gas ejected from the halo itself. The model has been proven flexible enough to describe the matter suppression in a large set of hydrodynamical simulations, both at the level of the matter power spectrum and bispectrum \citep{AricoEtal2021a}. It has also been successfully applied to data, fitting the shear signal measured by DES-Y3 \citep{ChenEtal2023, AricoEtal2023b}, fitting galaxy cluster properties \citep{GrandisEtal2024}, and jointly fitting the shear signal from DES-Y3, KiDS-1000 and HSC-DR1 \citep{Garcia-GarciaEtal2024}.

The emulator employed has an emulation accuracy of $1$-$2\%$ at $k=5~\ihMpc$. Apart from the cosmological parameters, it depends on 7 BCM parameters: $M_{\rm c}$ controls the typical halo mass at which half of the gas component has been expelled, and $\beta$ the slope of the dependence of the gas profile on halo mass; $\vartheta_{\rm inn}, M_{\rm inn}$, and $\vartheta_{\rm out}$ control the broken power low describing the virialised gas profile; $\eta$ controls the distance to which the gas is ejected from haloes; finally, $M_{1, z0, \mathrm{cen}}$ controls the characteristic mass of central galaxies at $z=0$.

This emulator is used in combination with another emulator of the \baccoemu{} family, providing predictions for the nonlinear boost of the matter power spectrum \citep{AnguloEtal2021}, applied to the emulated linear predictions of the matter power spectrum itself \citep{AricoEtal2021}. Also, in this case, the emulator is accurate at the $1$-$2\%$ level at $k=5~\ihMpc$.

\section{Simulations}\label{sec:sims}
\begin{figure*}
    \centering
    \includegraphics[width=\textwidth]{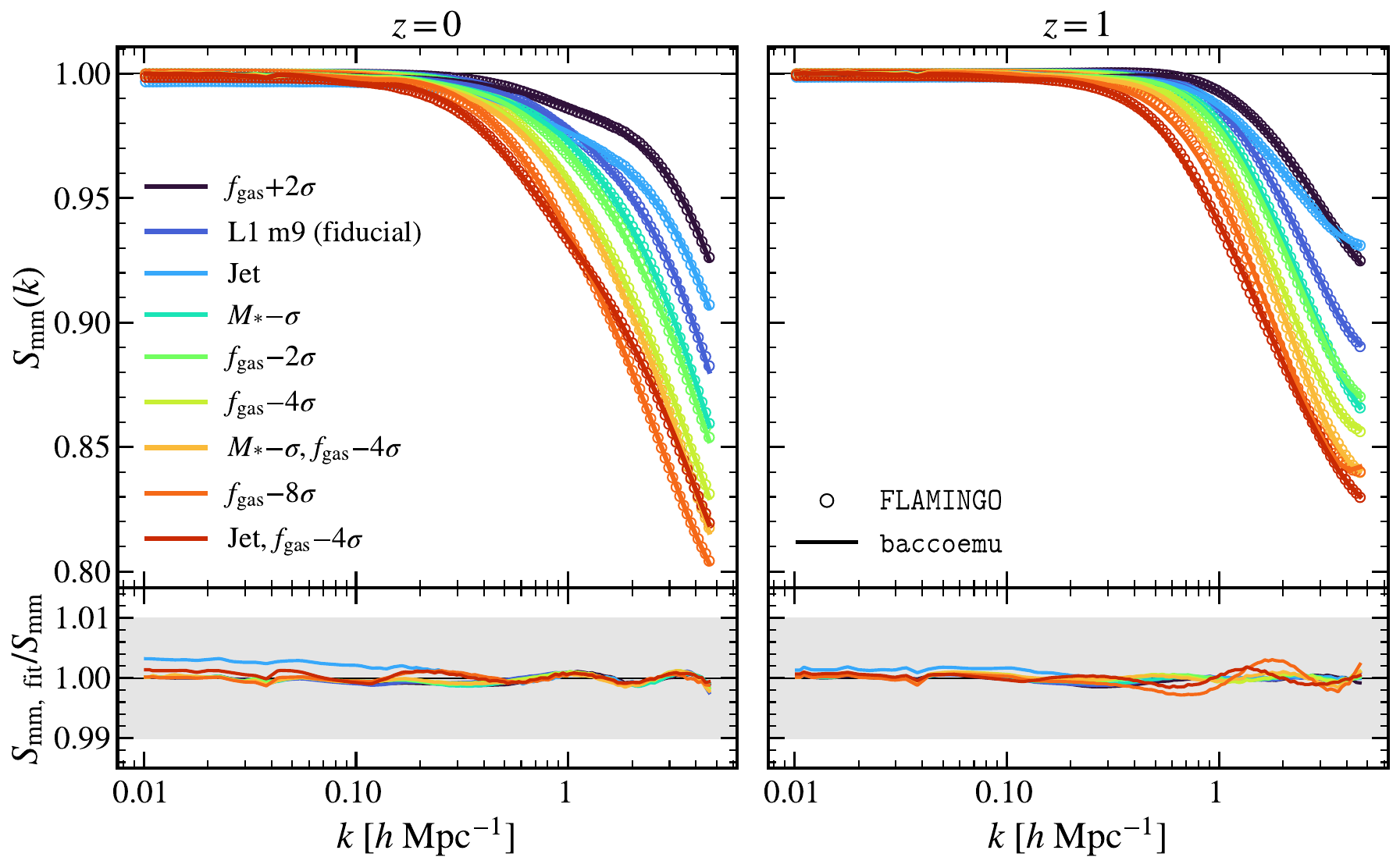}
    \caption{\textit{Open markers}: the baryon suppression $S_{\rm mm}(k) = P_{\rm mm, hydro}(k) / P_{\rm mm, dmo}(k)$ as measured for the 9 hydrodynamical simulations considered. \textit{Lines}: the predictions obtained with the \baccoemu{} emulator for the best fitting BCM parameters for each baryon case (obtained by fitting the measured suppressions from the simulations with \baccoemu{}). On the left we show redshift $z=0$ and on the right $z=1$. Lower panels show the ratio between the $S_{\rm mm}$ predictions obtained with our best fitting parameters and the $S_{\rm mm}$ from the simulations. Different baryonic feedback prescriptions are shown in order of increasing strength of the suppression at $k=1~\ihMpc$.}
    \label{fig:flamingo-Sk}
\end{figure*}
In this work, we use the state-of-the-art \flamingo{} simulations \citep{SchayeEtal2023,KugelEtal2023}. This is one of the largest hydrodynamical simulation suites, including baryon physics in different scenarios. The simulations were run with a version of the SWIFT code \citep{SchallerEtal2024}, in which the Smoothed Particle Hydrodynamics (SPH) is solved with the SPHENIX SPH scheme \citep{BorrowEtal2022}, particularly suited for galaxy formation simulations.

Specifically, we focus on the set of simulations sharing the same baseline cosmology (dubbed D3A), with $N_{\rm dm, part} = 1800^3$, an equal number of initial gas particles, and $N_{\rm \nu, part} = 1000^3$ neutrino particles, and with $L_{\rm box} = 861 ~ \hMpc$ (corresponding to $1000 ~ \mathrm{Mpc}$). The dark matter mass particles in these simulations have mass $M_{\rm DM} = 5.65 \times 10^9 \mathrm{M}_\odot$, while the initial baryonic particles have mass $M_{\rm g} = 1.07 \times 10^9 \mathrm{M}_\odot$. The D3A cosmology shared by these simulations is a flat cosmology with total matter density parameter $\Omega_{\rm m} = 0.306$, baryon density parameter $\Omega_{\rm b} = 0.0486$, and a neutrino component with total mass $M_\nu = 0.06 ~ \mathrm{eV}$. The power law index of primordial scalar perturbations is $n_{\rm s} = 0.967$ and the primordial power spectrum amplitude at a pivot scale $k_{\rm pivot} = 0.05 ~ \mathrm{Mpc}^{-1}$ is $A_{\rm s} = 2.099 \times 10^{-9}$. This corresponds to a standard deviation of linear perturbations (smoothed on spheres of radius $8 ~ \hMpc$) at $z=0$ of $\sigma_{8, \mathrm{tot}} = 0.807$ when considering perturbations in the total matter, or $\sigma_{8, \mathrm{cb}} = 0.811$ when considering only cold matter (CDM + baryons) perturbations. The Hubble parameter at redshift zero is $H_0 = 68.1 ~ \mathrm{km} ~ \mathrm{s}^{-1} ~ \mathrm{Mpc}^{-1}$. These parameters were chosen based on the Dark Energy Survey Year 3 (3x2point plus external constraints) analysis \citep{AbbottEtal2022}.

\subsection{Baryonic feedback models}
We consider nine baryon models from the \flamingo{} simulations, all sharing the same baseline cosmology. The different baryonic scenarios are summarised in Tab. \ref{tab:baryon-names}. All simulations include key astrophysical processes that are expected in the real universe, such as radiative cooling and heating, stellar and AGN feedback, chemical enrichment, and several processes affecting black hole spins and mergers \citep[see][for details]{SchayeEtal2023}. The fiducial model has been calibrated by setting the sub-grid parameters to reproduce the $z=0$ galaxy mass function and low-redshift galaxy cluster gas fractions. In contrast, other sub-grid parameters have been set based on numerical considerations \citep[see][for an extensive description of how each sub-grid parameter has been calibrated]{KugelEtal2023}. Four of the other models used, tagged "$f_{\rm gas} \pm N \sigma$", have been calibrated to reproduce modified cluster gas fractions with respect to the one assumed in the fiducial run. Similarly, another model, tagged "$M_*-\sigma$", has been calibrated to reproduce a modified stellar mass function, while "$M_*-\sigma, f_{\rm gas} - 4 \sigma$" reproduces both a different cluster gas fraction and stellar mass function with respect to the fiducial case. Finally, two of the models considered ("Jet" and "Jet, $f_{\rm gas} - 4 \sigma$") implement jet-like AGN feedback \citep{HuskoEtal2024} instead of thermal AGN feedback \citep{BoothSchaye2009}. 

\begin{table}
    \centering
    \begin{tabular}{|p{0.25\linewidth}|p{0.65\linewidth}|}
        \hline
        $f_{\rm gas}$+2$\sigma$                            & reproduce the cluster gas fraction shifted $2\sigma$ higher than the fiducial case; weaker feedback \\
        \hline
        L1 m9                                              & fiducial model \\
        \hline
        $f_{\rm gas}$$-$2$\sigma$                          & reproduce the cluster gas fraction shifted $2\sigma$ lower than the fiducial case; stronger feedback \\
        \hline
        $f_{\rm gas}$$-$4$\sigma$                          & reproduce the cluster gas fraction shifted $4\sigma$ lower than the fiducial case; stronger feedback \\
        \hline
        $f_{\rm gas}$$-$8$\sigma$                          & reproduce the cluster gas fraction shifted $8\sigma$ lower than the fiducial case; stronger feedback \\
        \hline
        $M_{*}$$-$$\sigma$                                 & reproduce the stellar mass function shifted $1\sigma$ lower than fiducial; stronger feedback\\
        \hline
        $M_{*}$$-$$\sigma$, $f_{\rm gas}$$-$4$\sigma$      & reproduce the stellar mass function shifted $1\sigma$ lower than the fiducial case and the cluster gas fraction shifted $4\sigma$ lower than the fiducial case; stronger feedback\\
        \hline
        Jet                                               & reproduce the same cluster gas fraction and stellar mass function as the fiducial case, but with jet-like AGN feedback\\
        \hline
        Jet, $f_{\rm gas}$$-$4$\sigma$                    & reproduce the cluster gas fraction shifted $4\sigma$ lower than the fiducial case with jet-like AGN feedback; stronger feedback\\
        \hline
    \end{tabular}
    \caption{Different available baryon scenarios considered in the \flamingo{} simulations with the same baseline cosmology.}
    \label{tab:baryon-names}
\end{table}

\begin{figure*}
    \centering
    \includegraphics[width=\textwidth]{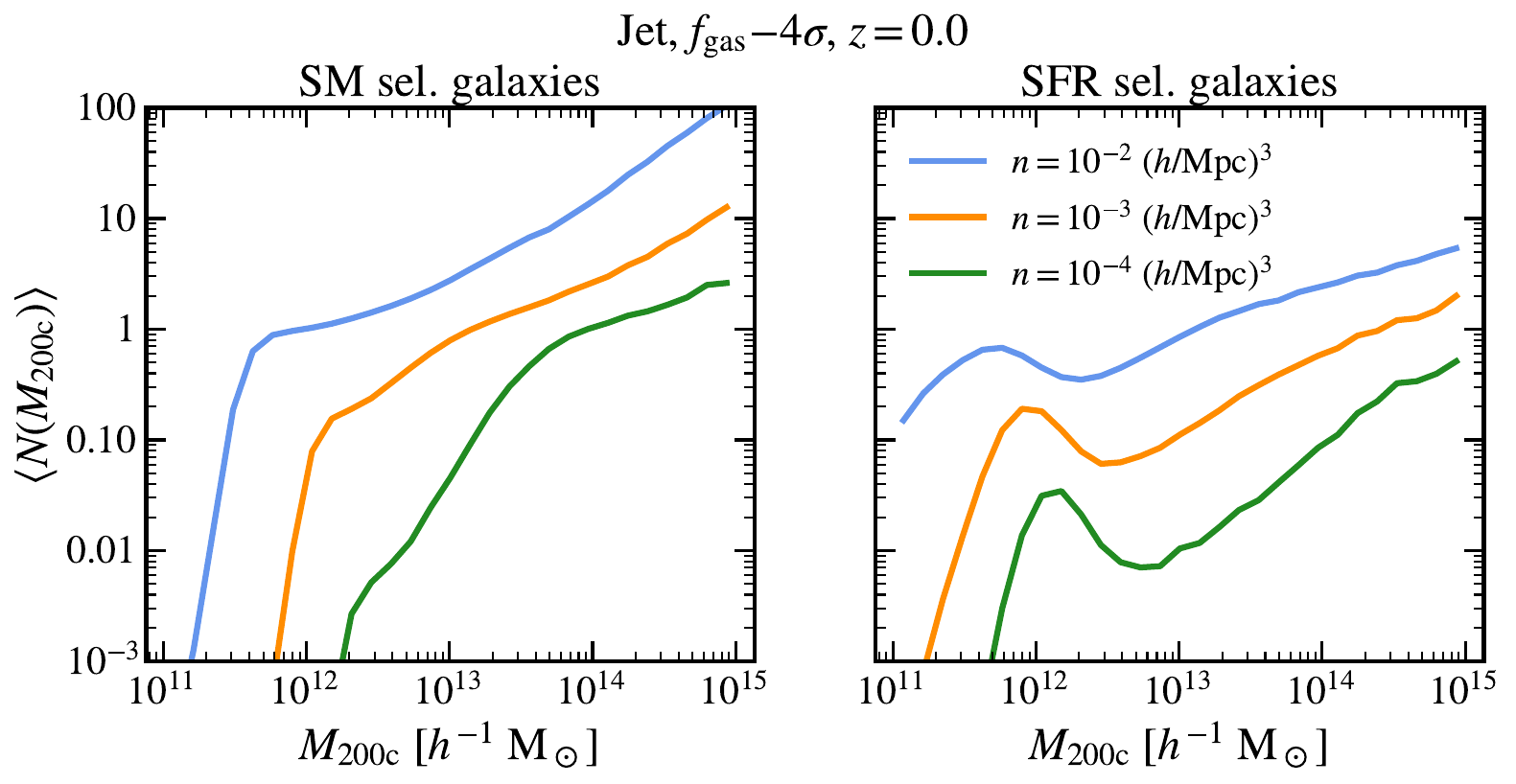}
    \caption{Example of HODs of galaxies in one of the baryonic models ("Jet, $f_{\rm gas}-4\sigma$") at $z=0$. On the left, galaxies are selected in order of decreasing Stellar Mass; on the right, in order of decreasing Star Formation Rate. Different colors correspond to different number densities, namely $n=\{10^{-2}, 10^{-3}, 10^{-4}\} \, h^3 \, \mathrm{Mpc}^{-3}$.}
    \label{fig:gal-hods}
\end{figure*}

Fig. \ref{fig:flamingo-Sk} shows the baryon-induced suppression of the matter power spectrum measured in the considered \flamingo{} models. This is shown until $k_{\rm max} = 5 ~ \ihMpc$, corresponding to the smallest scale included in the public version of the BCM emulator \baccoemu{} \citep{AricoEtal2020,AricoEtal2021b}.

The different baryon scenarios correspond to varying levels of suppression of the matter power spectrum, spanning suppressions from $\approx 7 \%$ to $\approx 20\%$ at $k=5 ~ \ihMpc$ and $z=0$, while suppressions are weaker at higher redshifts. 

Since we are interested in the scales covered by modern galaxy bias models (such as the hybrid Lagrangian bias expansion model), we will now focus on scales around $k = 1 ~ \ihMpc$. This corresponds to the smallest scale typically included in these types of analyses \citep[see][]{ZennaroEtal2023}. The model exhibiting the largest suppression at $k = 1 ~ \ihMpc$ is the one with jet-like AGN feedback, dubbed "Jet, $f_{\rm gas}-4\sigma$"; for this reason, whenever we will focus on a single model, we will use this as our reference case.

\begin{figure*}
    \centering
    \includegraphics[width=\linewidth]{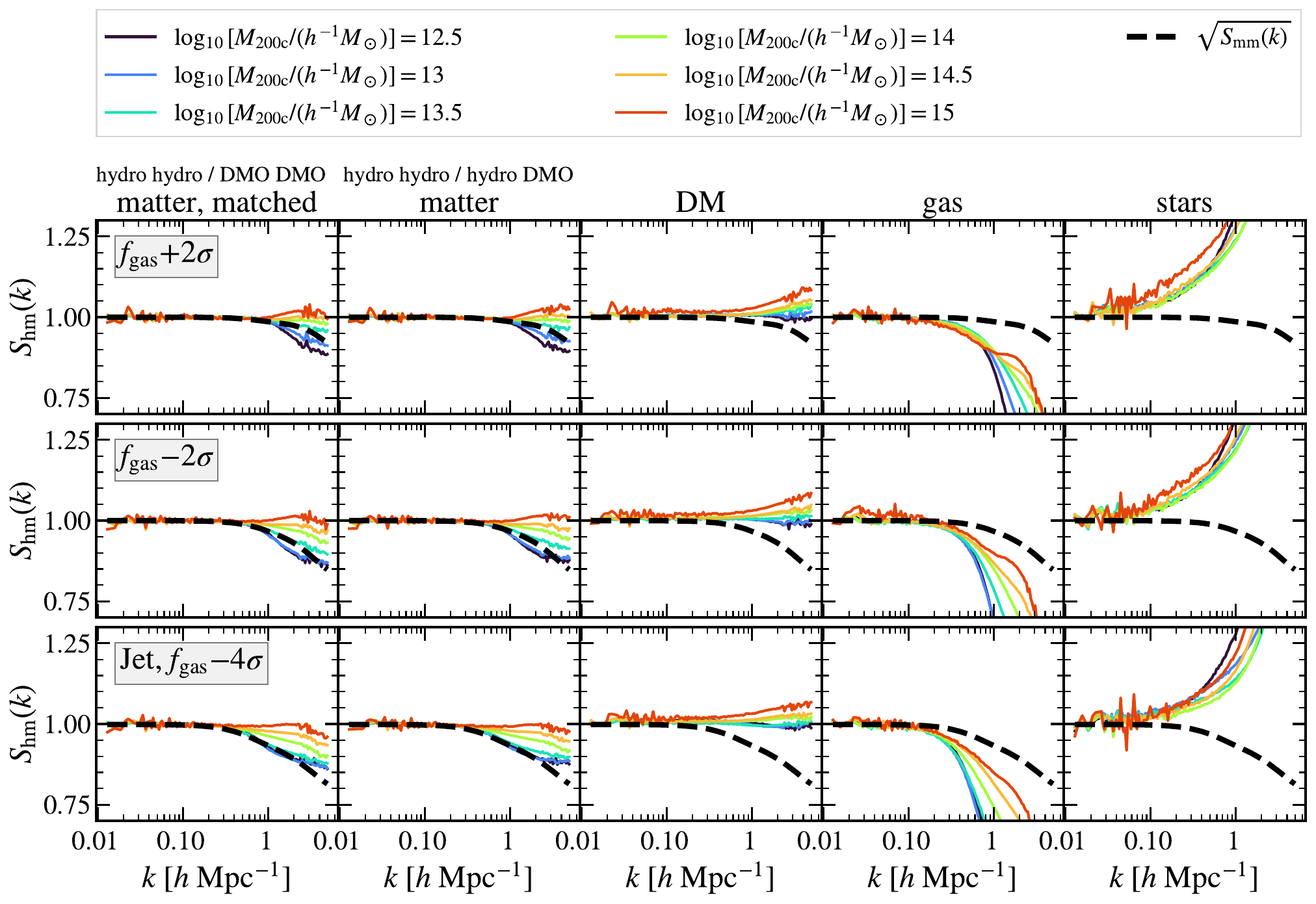}
    \caption{Comparison of the suppression $S_{\rm hm}(k) = P_{\rm hm, hydro} / P_{\rm hm, dmo}$ for central haloes in three hydrodynamical simulations (different rows) and the respective DMO at $z=0$. Haloes are split into mass bins (different line colours). First column: the ratio between the halo-matter cross-power spectrum from the hydrodynamic and the DMO simulations; in this case, haloes are matched between the hydrodynamical and DMO simulations. From the second column and to the right: the ratio of the cross-power spectrum of haloes and matter in the hydrodynamical simulation and the cross-power spectrum of the same haloes (from the hydrodynamical simulation) and matter from the DMO simulation; the matter component of the hydrodynamical simulation is either the total matter (second column), only the dark matter (third column), only the gas (fourth column), or only the stellar component (last column). In all cases, the suppression $\sqrt{S_{\rm mm}(k)}$ inferred from the matter fields (from the hydrodynamical versus DMO simulations) is shown for reference as a black dashed line. Haloes of mass $10^{13.5}$ to $10^{14} ~ h^{-1} \mathrm{M}_{\odot}$ (which contribute the most to the total matter power spectrum on the scales of interest here) exhibit suppressions well approximated by $\sqrt{S_{\rm mm}}$, while lower and higher mass haloes exhibit (respectively) stronger and weaker suppressions, due to the different behaviour of their gas and star components and the differences in the consequent DM adiabatic relaxation.}
    \label{fig:subhaloes}
\end{figure*}

\subsection{Haloes}\label{sec:subhaloes}
Haloes and subhaloes in the \flamingo{} simulations are identified using the \texttt{VELOCIraptor} algorithm \citep{ElahiEtal2019}, and the galaxy properties in each halo are obtained with a \flamingo{}-specific tool called SOAP, Spherical Overdensity and Aperture processor \citep[see][for details]{SchayeEtal2023}. We will use galaxy properties computed by SOAP in apertures of 50 kpc, excluding unbound particles. 

For the first part of our results (\S \ref{sec:halomassbins}) we consider central haloes, focusing on overdensities crossing the threshold set by 200 times the critical density in the given cosmology, and consider the corresponding halo mass $M_{200\mathrm{c}}$.

To avoid resolution effects, we select haloes (both in the hydrodynamical and the corresponding Dark Matter Only, or DMO, simulations) at fixed abundance $n=10^{-2} ~ h^3 ~ \mathrm{Mpc}^{-3}$. For the DMO simulation, this corresponds to a minimum spherical overdensity (SO) mass of $M_{200\mathrm{c}} = 7.52 \times 10^{11} ~ h^{-1} ~ \mathrm{M}_\odot$, corresponding to approximately 160 DM particles.

Subsequently, we split the halo sample into different mass bins spanning $\log_{10}[M_{200\mathrm{c}} / h^{-1} M_{\odot}] \in [12.25, 15.25]$, each bin having a 0.5 dex width. 

\subsection{Galaxy samples}\label{sec:galaxies}
In each hydrodynamical simulation, we select galaxies based on two different criteria. In one case, we select galaxies with the largest stellar masses, roughly corresponding to luminous, redder galaxies, most abundantly found in galaxy clusters and high-density regions. Throughout this work, we will refer to such galaxies as stellar mass (SM) selected. In the other case, we rank galaxies based on their star formation rate (and we will thus refer to these samples as SFR selected). They more likely populate filaments and low-density regions, are mostly central galaxies of their host halo, and are bluer and characterised by multiple emission lines. 

Moreover, within each type of galaxy selection (and in each baryonic feedback model), we choose our galaxy samples to have three different number densities ($n = 10^{-2}, 10^{-3}, 10^{-4} ~ h^3 ~ \mathrm{Mpc}^{-3}$). These number densities are qualitatively (i.e. not accounting for a realistic redshift distribution) similar to those expected for Stage-IV surveys, with the range $10^{-4}-10^{-3} ~ h^3 ~ \mathrm{Mpc}^{-3}$ being similar to the expected DESI and Euclid spectroscopic samples, and $10^{-3}-10^{-2} ~ h^3 ~ \mathrm{Mpc}^{-3}$ being in the range of LSST and Euclid photometric samples \citep{EuclidPresentation2024}. This ensures that host haloes of significantly different masses are relevant in each sample, thus guaranteeing that our results are general. As an example, in Fig. \ref{fig:gal-hods} we show the Halo Occupation Distribution (HOD), that is, the mean number of galaxies that occupy haloes as a function of halo mass, for one baryonic model ("JETS, $f_{\rm gas} - 4 \sigma$", which is the strong jets AGN model) at $z=0$ for the different number densities, both for SM selected and SFR selected galaxies. The other galaxy samples populate haloes of different mean masses and exhibit a diverse satellite distribution.

We note that even sparser galaxy samples are possible depending on the survey geometry and selection criteria. However, for such samples, the level of shot noise ($P_{\rm sn} > 10^4 ~ h^{-3} ~ \mathrm{Mpc}^3$) becomes dominant on rather large scales at typical survey redshifts, making the modelling of baryon effects rather unimportant.

Finally, these choices for selecting haloes and galaxies could depend on the mass resolution of the simulations used. In appendix \ref{appendix:res}, we investigate how the simulation resolution affects the halo and galaxy abundances (Fig. \ref{fig:hod_hr}). We find that, while the standard resolution adopted in this work is not enough to recover small mass haloes, thus introducing modifications of the HODs, it is sufficient to select galaxy samples that populate haloes of very different masses and with different biasing properties with respect to the underlying dark matter. We will discuss later how this affects the results of our analysis.

\section{Validation with the \flamingo{} simulations}\label{sec:validation-with-sims}
\begin{figure*}
    \centering
    \includegraphics[width=0.93\textwidth]{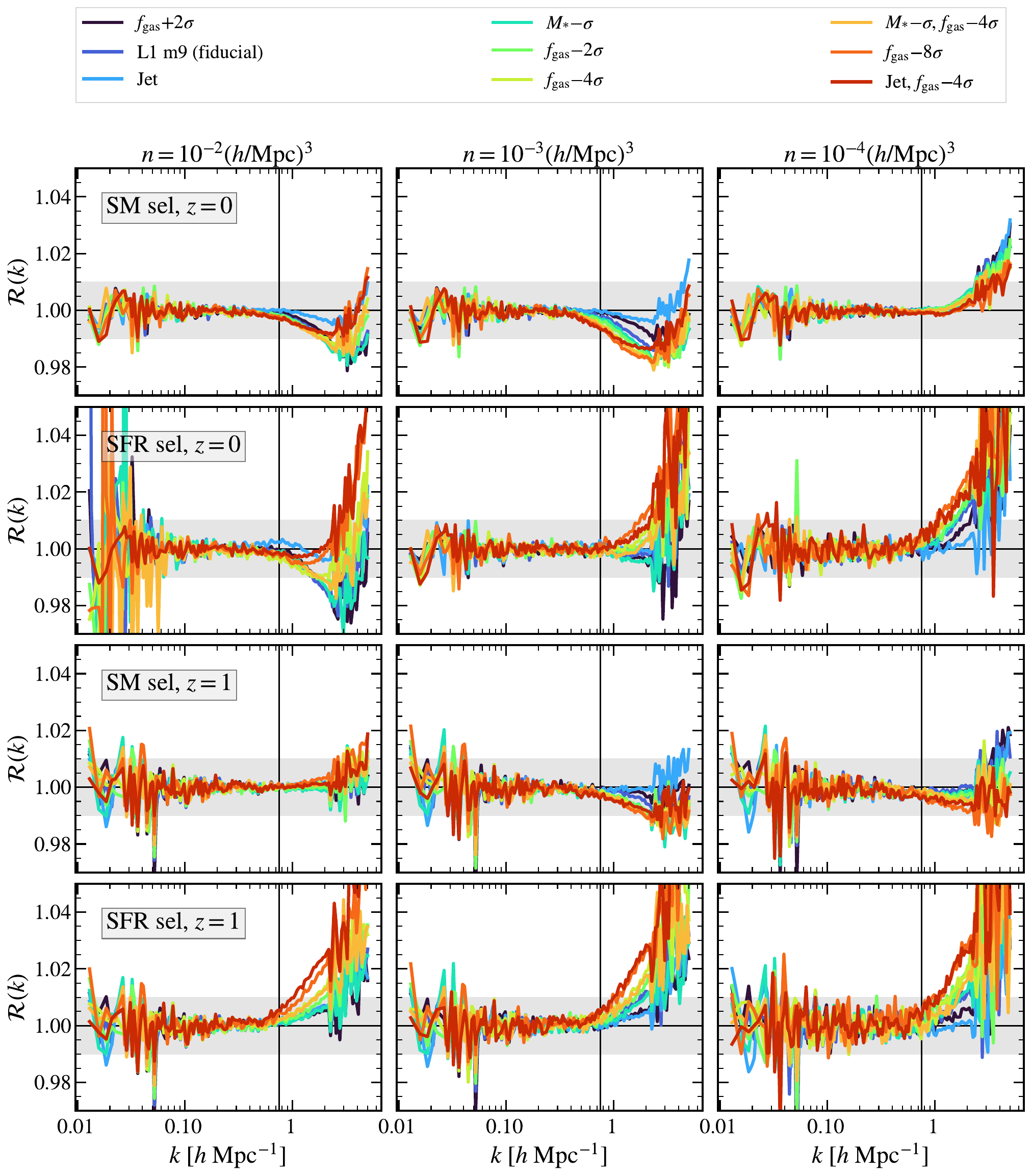}
    \caption{The ratio $\mathcal{R}(k) = P_{\rm gm, hydro}(k) / [P_{\rm gm, dmo}(k) \sqrt{S_{\rm mm}(k)}]$ obtained with spectra directly measured in the simulations considered. The three columns correspond to samples with number density $n = \{10^{-2}, 10^{-3}, 10^{-4} \} ~ h^3 ~\mathrm{Mpc}^{-3}$. The different rows correspond to SM-selected and SFR-selected galaxies at $z=0$ (upper two rows) and $z=1$ (lower two rows). The vertical black line marks $k=0.7~\ihMpc$, i.e. the limiting scale of the hybrid Lagrangian bias expansion model. In all cases, the approximation proposed in this work is 1\% accurate on the range of scales allowed by the hybrid galaxy model.}
    \label{fig:sqrtSk-all}
\end{figure*}
\subsection{Haloes of different masses}\label{sec:halomassbins}
Before moving to galaxies, we investigate the effect of baryons on the halo-matter cross-power spectrum. We consider the halo populations defined in Section \ref{sec:subhaloes}; specifically, we consider central haloes at $z=0$, with fixed abundance ($n=10^{-2} ~ h^3 \mathrm{Mpc}^{-3}$), among which we select mass bins of width 0.5 dex.

We consider three baryonic physics scenarios: the one with the smallest suppression at $k=1 ~ \ihMpc$ ($f_{\rm gas}+2\sigma$, i.e. weak feedback on cluster scales), the one with the largest suppression at the same scale (Jet, $f_{\rm gas}-4\sigma$, i.e. strong jets AGN), and an intermediate case ($f_{\rm gas} - 2\sigma$, i.e. strong AGN), according to the suppressions presented in Fig. \ref{fig:flamingo-Sk}.

In Fig. \ref{fig:subhaloes}, we show two quantities: the ratio between the halo-matter cross-power spectrum in the hydrodynamical simulation and the same quantity in the DMO simulation, using haloes \textit{matched} between the two simulations \footnote{Matched haloes between the hydro and DMO runs are found by requiring the difference of the position of the halo centres not to exceed 3 times $R_{200\mathrm{c}}$ and their logarithmic mass difference (computed with $M_{200\mathrm{c}}$) not to exceed 1. Details of the implementation of the matching algorithm can be found in Ondaro-Mallea et al, in prep} (and using the halo mass from the DMO simulation to define the mass bins). The ratio between the cross-power spectrum of haloes from the hydrodynamical simulation and the matter from the hydrodynamical simulation is divided by the same quantity computed with the same haloes from the hydrodynamical simulation but with the matter from the DMO simulation. For the last case, we consider the matter of the hydrodynamical simulation in its totality, or we split it into its components, namely DM, gas, and stars.

In the first column of Fig. \ref{fig:subhaloes}, we consider central haloes (identified in the simulation as not being substructures of larger haloes) matched between the hydrodynamical and DMO simulations. We see that baryonic physics does induce differences in the halo-matter cross-power spectrum. Specifically, we find the suppression to be typically larger for low-mass haloes than for the matter field (i.e., our proposed model); the opposite happens for high-mass haloes.

We now investigate this mass dependence by considering the different cross-correlations of the same halo sample with the different matter components, considering the total matter, dark matter (DM), gas, and stars from the hydrodynamical simulation.

First, we consider haloes from the hydrodynamical simulation and cross-correlate them with the total matter from the same simulation and the corresponding DMO simulation (second column of Fig. \ref{fig:subhaloes}). Once again, we find a trend with the suppression for low-mass haloes falling below to the matter case (stronger suppression than for the matter), and, for high-mass haloes, falling above the suppression inferred from matter (weaker suppression than for the matter). For haloes between $10^{13.5}$ and $10^{14} ~ h^{-1} ~ \mathrm{M}_\odot$, the suppression inferred from the matter component is a good approximation for the halo-matter cross-power spectrum. This is likely related to haloes of this mass being the most dominant contribution to the matter power spectrum \citep[e.g.][]{VanDaalenSchaye2015,DebackereEtal2020,SalcidoEtal2023,VanLoonVanDaalen2024}. This is in agreement with what was found for the matched haloes.

Next, we repeat this experiment but consider the DM, gas and star components separately. The cross-correlation with the star component (last column of Fig. \ref{fig:subhaloes}) shows an increase of power (with respect to the DMO case) due to galaxy formation processes, especially the formation of the central galaxy, with only a weak dependence on halo mass; this contribution is by far subdominant in the total matter.  The cross-correlation with the gas component (fourth column of Fig. \ref{fig:subhaloes}), on the other hand, shows a significant lack of power with respect to the DMO case due to the displacement of gas in the halo, with this suppression being stronger for lower halo masses; this depends on the fact that smaller mass haloes are more strongly affected by feedback processes and are more likely to lose their gas. Finally, the cross-correlation with the DM component (third column of Fig. \ref{fig:subhaloes}) shows an increase of power with respect to the DMO case, reflecting the contraction of the DM induced by the formation of the central galaxy; this effect is more prominent for high-mass haloes and almost negligible for small-mass haloes.

In this picture, low-mass haloes expel a vast amount of gas while forming galaxies that do not substantially modify the DM distribution at their centre, explaining why, in this case, the halo-matter cross-correlation exhibits a stronger suppression than the one inferred from the matter component alone.

On the other hand, high-mass haloes still exhibit a significant power suppression due to gas displacement, but it is less important than for haloes of small mass; at the same time, the formation of the galactic component, in this case, imprints an increase of power, on small scales in the DM distribution. This results in their halo-matter cross-power spectrum being less suppressed than the matter power spectrum\footnote{We repeated this exercise for subhaloes. For matched subhaloes, the number of objects in each mass bin is too small to draw robust conclusions. For unmatched subhaloes, we find the same trend as for central haloes.}.

\subsection{Galaxy samples}\label{sec:Ratios}
In Fig. \ref{fig:sqrtSk-all}, we show the performance of applying the $\sqrt{S_{\rm mm}(k)}$ correction to galaxy samples. 

Specifically, we define the metric
\begin{equation}
    \mathcal{R}(k) = \dfrac{P_{\rm gm, hydro}}{P_{\rm gm, dmo} \sqrt{S_{\rm mm}(k)}}~,
\end{equation}
where the galaxy sample is fixed (and defined in the hydrodynamical simulation) and is cross-correlated with the matter from the hydrodynamical simulation in the numerator and of the DMO simulation in the denominator; $S_{\rm mm}(k)$ is the baryonic suppression inferred from the matter auto power spectra of the hydrodynamical and DMO simulations. This metric quantifies the accuracy (as a function of scale) of describing the galaxy matter cross-power spectrum in the presence of baryons as the same quantity predicted for a baryonless case corrected by a suppression inferred from the matter component.

This approximate baryon description is 1\% accurate down to $k \approx 1 \ihMpc$ for both SM- and SFR-selected galaxies. This is accurate enough for practical applications since we are interested in describing scales $k \leq 0.7 \ihMpc$, where hybrid Lagrangian bias expansion models can be employed. 

On smaller scales, this approximation breaks down, suggesting that the effects of baryons on the galaxy and matter fields are not separable, as also found by \cite{VanDaalenEtal2014}.

For example, the densest sample of SM-selected galaxies at $z=0$ (top left panel of Fig. \ref{fig:sqrtSk-all}) shows that when the $\sqrt{S_{\rm mm}(k)}$ approximation breaks down, it predicts a smaller suppression than the one exhibited by the galaxy matter cross-correlation -- resulting in our metric $\mathcal{R}$ falling below 1. This is consistent with what we find in Fig. \ref{fig:subhaloes}, where smaller mass haloes (the ones contributing the most to this galaxy sample, as shown in Fig. \ref{fig:gal-hods}) are affected by stronger baryonic suppression than the entire matter field. On the other hand, the sparsest sample of SM galaxies at $z=0$ (top right panel of Fig. \ref{fig:sqrtSk-all}) shows a weaker baryon suppression, consistently with what we find for high mass haloes in Fig. \ref{fig:subhaloes} -- high mass haloes being the ones most commonly hosting these galaxies as shown in Fig. \ref{fig:gal-hods}.

Similarly, the densest SFR-selected sample from the "Jet, $f_{\rm gas}-4\sigma$" simulation at $z=0$ shows that the measured cross-power spectrum is just slightly more suppressed than what the matter approximation predicts (less than a few permille) at $k=1~\ihMpc$. This is compatible with this sample being dominated by galaxies preferentially residing in low mass haloes (Fig. \ref{fig:gal-hods}), and these exhibiting a very slightly stronger suppression than the reference $\sqrt{S_{\rm mm}}$ (Fig. \ref{fig:subhaloes}). On the other hand, the sparsest SFR selected sample at $z=0$ shows departures above one at $k=1~\ihMpc$. From Fig. \ref{fig:gal-hods}, we can see that this sample is typically hosted in more massive haloes, which we see in Fig. \ref{fig:subhaloes} usually exhibit weaker suppressions than $\sqrt{S_{\rm mm}}$. 

Finally, we remark again that these results can, in principle, depend on the mass resolution of the simulations adopted. However, in appendix \ref{appendix:res}, particularly in Fig. \ref{fig:RR_HR}, we show that the results presented in this section are robust against using a simulation with higher mass resolution. 

\section{Bayesian analysis}\label{sec:bayesian}
\subsection{Likelihood}\label{sec:likelihood}
To assess the impact of baryons on the galaxy-matter cross-power spectrum and the subsequent inference of cosmological and galaxy bias parameters, we perform Bayesian analyses using the \multinest{} nested sampler \citep{FerozHobsonBridges2008}, with its python wrapper \texttt{pymultinest} \citep{BuchnerEtal2014}. We use $N_{\rm live} = 400$ live points and stop exploring posteriors when a 0.1 dex tolerance on the estimate of the log-evidence is reached.

We build our data vectors either as $\lbrace P_{\rm gg}(k), P_{\rm gm}(k) \rbrace$ or $\lbrace P_{\rm gg}(k), P_{\rm gm}(k), P_{\rm mm}(k) \rbrace$, depending on the case. Here $P_{\rm gg}(k)$ is the galaxy auto-power spectrum, $P_{\rm gm}$ is the galaxy-matter cross-power spectrum, and $P_{\rm mm}$ is the matter auto-power spectrum. For $P_{\rm gm}$ and $P_{\rm mm}$, the matter component can come from the DMO or hydrodynamical simulation.

All power spectra are computed by assigning mass to a regular grid of $N_{\rm grid} = 1024$ with a Cloud-in-Cell (CIC) assignment scheme. These meshes are used both directly and after folding them 8 times per direction to access smaller scales with the same $N_{\rm grid}$ \citep{JenkinsEtal1998,ColombiEtal2009,AricoEtal2021a}. We also deconvolve our meshes to correct for the window function introduced by the mass assignment scheme, and we use interlacing to reduce aliasing \citep{SefusattiEtal2016}.
 
Errors on the different power spectra are computed assuming cosmic variance for a cosmic volume corresponding to the simulation size and a stochastic component. Therefore, each block of the final covariance matrix is assumed to be diagonal, and we compute it in the Gaussian approximation:

\begin{equation}
    \sigma^2 [P_{ij}(k)] = \dfrac{2}{N_k} \left[ P_{ij}(k) + \dfrac{\delta_{ij}^{\rm D}}{\bar{n}} \right]^2,
\end{equation}

\noindent with $N_k = [V/(2 \pi)^3] 4 \pi k^3 \mathrm{d} \ln k$ the number of modes falling in each $k$-bin, and $\delta_{ij}^{\rm D}$ the Dirac delta function adding the shot noise contribution to auto-power spectra, but not to cross-power spectra. We assume no cross covariance between $P_{\rm gg}$, $P_{\rm gm}$ and $P_{\rm mm}$.

The resulting covariance matrix is not an accurate representation of the actual data covariance -- even for simulated data. First, the small-scale power spectrum is expected to exhibit cross-covariance between $k$-bins as an effect of non-linear gravitational evolution; secondly, galaxy and matter are not decoupled; and, finally, the \flamingo{} simulations are set up with fixed amplitudes at the initial conditions for modes $(k/L)^2 < 1025$, corresponding, for the simulations considered here with $L=1000~\mathrm{Mpc}$ and $h=0.681$, to $k < 0.05 ~ \ihMpc$. Therefore, our covariance represents a lower limit of the error budget on small scales and is not expected to bias our results but to modify the size of our contours. In this respect, \cite{HouEtal2022} have shown that analytical Gaussian covariances are reasonable enough when dealing with data from simulations with periodic boundary boxes, while they perform poorly when a complex window function should be considered. For the sake of the problem at hand, therefore, it is fundamental that we treat data errors consistently, albeit not realistically. 

To perform our Bayesian analysis, we define a Gaussian Likelihood function:

\begin{equation}
    \ln \mathcal{L} = - \dfrac{1}{2} \left( \boldsymbol{d} - \boldsymbol{t} \right) C^{-1} \left( \boldsymbol{d} - \boldsymbol{t} \right)^\mathrm{T},
\end{equation}
modulo an additional constant, where $\boldsymbol{d}$ is our data vector, $\boldsymbol{t}$ is the theory vector, and $C$ the covariance matrix.

When fitting the full data vector (with galaxy auto-power spectrum, galaxy-matter cross-spectrum, and matter auto-power spectrum) with the full model (with hybrid Lagrangian galaxy bias, BCM and nonlinear matter power spectrum), we deal with 14 free parameters, reported in Tab. \ref{tab:priors}, along with the priors we assume for each of them. In the different cases considered, we will always state which parameters are left free and which are kept fixed.

\begin{table}
    \centering
    \begin{tabular}{|l|c|}
    \hline
    Parameter name & Prior \\
    \hline
       $b_{1}$  & $\mathcal{U}(-1, 3)$ \\
       $b_{2}$  & $\mathcal{U}(-3, 3)$ \\
       $b_{s^2}$  & $\mathcal{U}(-10, 10)$ \\
       $b_{\nabla^2\delta} ~ [h^{-2} ~ \mathrm{Mpc}^2]$ & $\mathcal{U}(-10, 10)$ \\
       $A_{\rm sn}$  & $\mathcal{U}(0, 3)$ \\
    \hline
       $\log_{10}[M_{\rm c} / (h^{-1} \mathrm{M}_\odot)]$ & $\mathcal{U}(9, 15)$ \\
       $\log_{10} \eta$ & $\mathcal{U}(-0.7, 0.7)$ \\
       $\log_{10} \beta$ & $\mathcal{U}(-1, 0.7)$ \\
       $\log_{10}[M_{1, z0, \mathrm{cen}} / (h^{-1} \mathrm{M}_\odot)]$ & $\mathcal{U}(9, 13)$ \\
       $\log_{10}\vartheta_{\rm inn}$ & $\mathcal{U}(-2, -0.523)$ \\
       $\log_{10}[M_{\rm inn} / (h^{-1} \mathrm{M}_\odot)]$ & $\mathcal{U}(9, 13.5)$ \\
       $\log_{10} \vartheta_{\rm out}$ & $\mathcal{U}(0, 0.48)$ \\
    \hline
       $\omega_{\rm cb} \equiv (\Omega_{\rm cdm} + \Omega_{\rm b}) h^2$ & $\mathcal{U}(0.107, 0.19)$ \\
       $\sigma_{8,\mathrm{cb}}$ & $\mathcal{U}(0.73, 0.9)$ \\
    \hline
       
    \end{tabular}
    \caption{The full set of free parameters in our analysis and their corresponding priors. We indicate with $\mathcal{U}(x, y)$ a uniform distribution defined in the interval $[x, y]$.}
    \label{tab:priors}
\end{table}

\begin{figure}
    \centering
    \includegraphics[width=\linewidth]{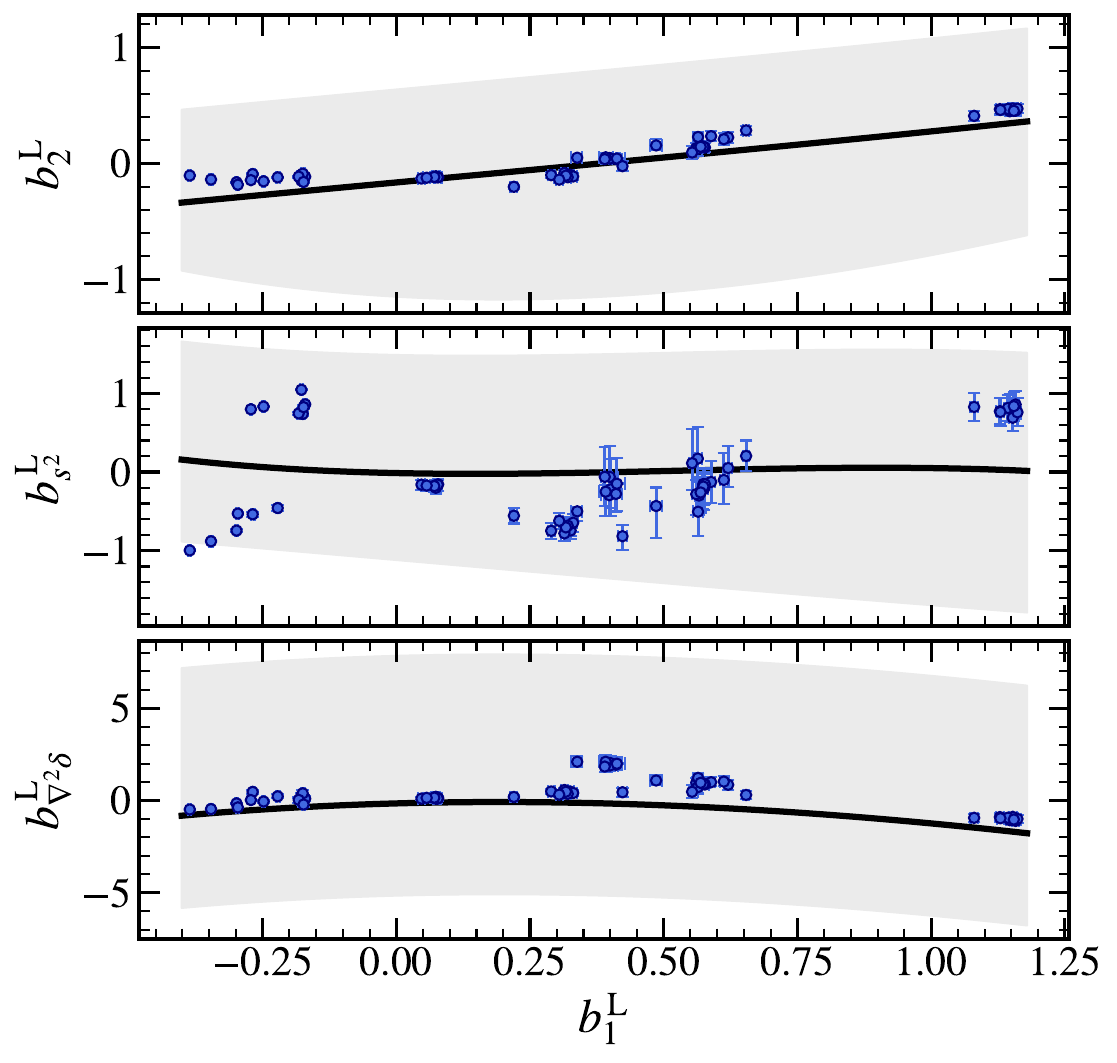}
    \caption{Coevolutions relations from fits of $P_{\rm gg}$ and $P_{\rm gm}$, where $P_{\rm gm}$ is computed by cross-correlating the galaxy distribution of the hydrodynamical simulations with the matter field of the corresponding DMO simulation. All baryon models at $z=\{0, 1\}$, all galaxy selection criteria (SM and SFR), and number densities $n_{\rm g} = \{10^{-2}, 10^{-3}\} ~ h^3 ~ \mathrm{Mpc}^{-3}$ are reported. The cross-correlation with the matter from the DMO simulations allows us to find the reference values of these bias parameters. Black solid lines and the grey shaded areas represent the coevolution relations and allowed parameter space from \cite{ZennaroEtal2022}.}
    \label{fig:coevolution relations}
\end{figure}

\begin{figure*}
    \centering
    \includegraphics[width=\textwidth]{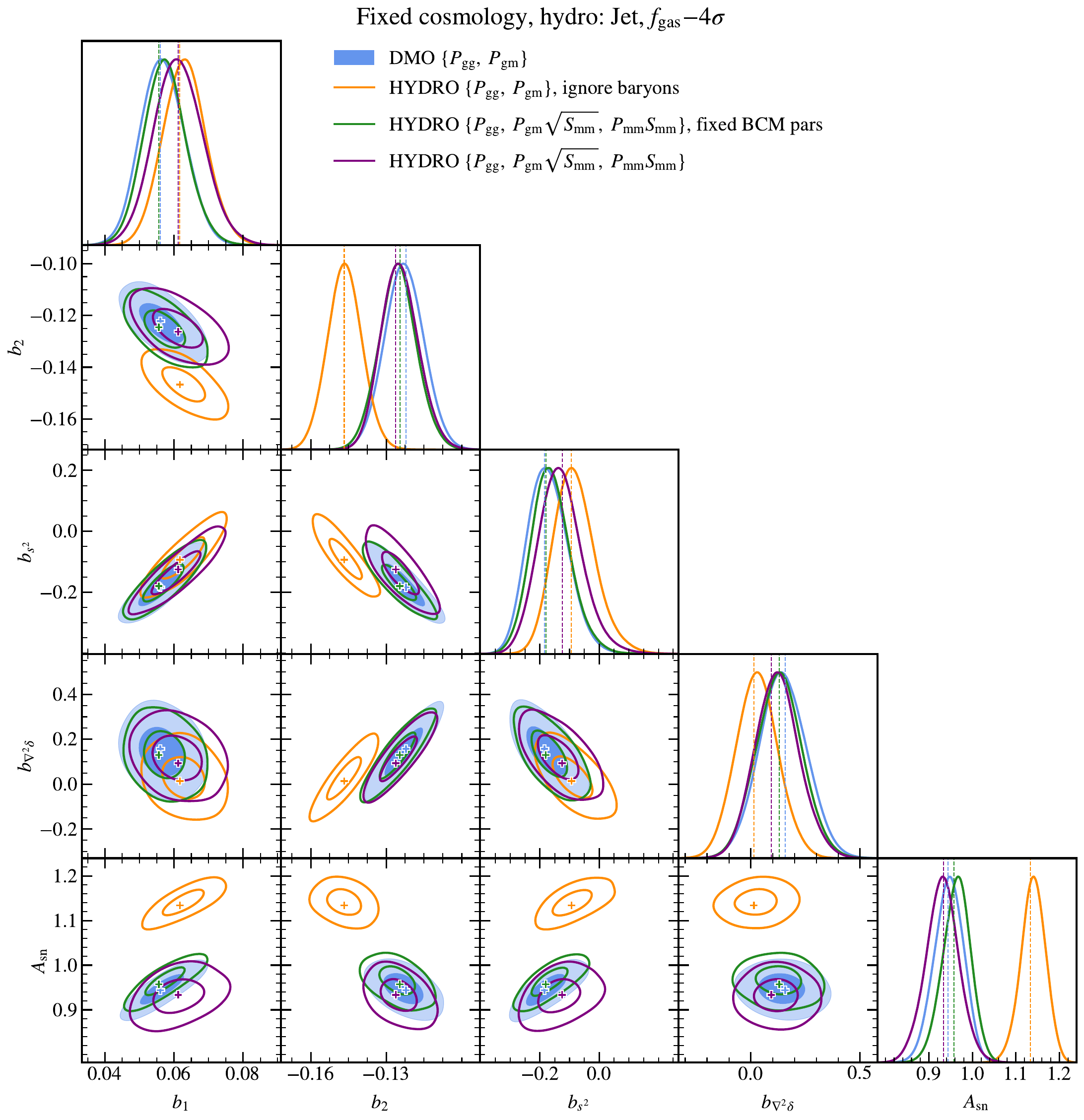}
    \caption{The 1- and 2-$\sigma$ contours of the posterior distribution obtained by fitting the clustering data ($P_{\rm gg}$ and $P_{\rm gm}$) at fixed cosmology for the "Jet, $f_{\rm gas} - 4\sigma$" simulation at $z=0$, with SM-selected galaxies with number density $n=10^{-2} ~ h^3 ~ \mathrm{Mpc}^{-3}$. Plus signs (in the 2D contours) and dotted lines (in the 1D distributions) mark the position of the best-fitting parameters.  Different colours correspond to different cases: \textit{blue} corresponds to the reference values of the bias parameters, where the data come from galaxies (from the hydrodynamical simulation) correlated with matter from the DMO simulation, and the model does not include baryons; in \textit{orange}, galaxies in the data vector are cross-correlated with the matter of the hydrodynamical simulation, but the model does not account for baryons; baryons are present in the data also in \textit{green}, and are accounted for in the model using the $\sqrt{S_{\rm mm}}$ approximation proposed in this work, with BCM parameters fixed to their best fitting values obtained from fitting the measured $S_{\rm mm}$; finally, in \textit{purple}, also the BCM parameters are left free to vary.}
    \label{fig:fixedcosmo}
\end{figure*}

\paragraph{A note on scales.} Whenever dealing with galaxy auto- and cross-spectra, we perform our analyses including scales up to $k_{\rm max} = 0.7 ~ \ihMpc$, or until the power spectrum signal falls below 1.5 times the shot noise level, whichever comes first. We note that for samples at $z=0$ and number density $n_{\rm g}=10^{-2} ~ h^3 ~ \mathrm{Mpc}^{-3}$ for all selection criteria, we are always able to reach $k_{\rm max} = 0.7 ~ \ihMpc$. We do not extend our analyses to scales smaller than this since the hybrid Lagrangian model considered (already among the models able to describe galaxy clustering to the smallest possible scales) is no longer valid. Whenever, instead, we consider the matter power spectrum (with or without baryons), we include scales down to $k_{\rm max} = 5 ~ \ihMpc$, which is the smallest scale described by the public \baccoemu{} emulator for both matter power spectra and baryons suppression.

\paragraph{A note on $\chi^2$ values.} We do not quote the $\chi^2$ (or other similar statistics) for our best-fitting parameters since, in our case, such numbers hold little meaning. Specifically, we approximate our power spectrum errors as drawn from a Gaussian distribution corresponding to the simulation volume, with no cross-covariance between different modes. However, these simulations have fixed amplitude initial conditions on scales $k < 0.05 ~ \ihMpc$, and random amplitude initial conditions on smaller scales. This makes it difficult to predict the correct covariance matrix associated with data measured in such simulations. One way to do so would be to produce a large set of mock data (with fast simulation codes) sharing the same initial conditions set up as the \flamingo{} simulations but representing different random realisations of the phase distribution (and initial amplitudes for small scales). However, having fully realistic error bars is not within the scope of this work, and our Gaussian covariance is sufficient, as discussed above, to not bias our results. Our conclusions, therefore,  can be drawn by comparing the shifts in our posteriors obtained employing consistent covariance matrices among the different cases rather than the specific size of the contours inferred.

\subsection{Matter power spectrum suppression}
We first obtain our reference values for the BCM parameters for the different hydrodynamical simulations considered. We do so by fitting the suppression $S_{\rm mm}(k) \equiv P_{\rm mm, hydro}(k) / P_{\rm mm, dmo}(k)$ with the model obtained from the BCM emulator \baccoemu{} down to scale $k_{\rm amx} = 5 ~ \ihMpc$. 
The best-fitting parameters of the BCM model found for the different hydrodynamical simulations are reported in Tab. \ref{tab:Smmz0-bf} (for $z=0$) and Tab. \ref{tab:Smmz1-bf} (for $z=1$).

In Fig. \ref{fig:flamingo-Sk}, we show that the BCM model, evaluated with the best fitting parameters corresponding to each baryonic scenario, provides a sub-percent description of the suppression measured directly from the simulations.

\subsection{Reference bias values}
As a first step, we investigate the values of the bias parameters expected for our samples of galaxies. The hybrid Lagrangian bias expansion model is meant to capture the effect of baryons on galaxy clustering, mainly through the freedom guaranteed by the Laplacian term $b_{\nabla^2\delta}$ \citep{DesjacquesJeongSchmidt2016,ModiChenWhite2020}. However, in the galaxy-matter cross-correlation, the model itself is not designed to capture the effect of baryons introduced by the matter component, especially not in a way consistent with the values of the bias parameters from the galaxy auto power spectrum. For this reason, our reference values of the bias parameters are inferred by fitting the auto power spectrum of galaxies from all the different hydrodynamical simulations (selected either by SM of SFR and with the number densities and redshifts presented in Sec. \ref{sec:galaxies}), together with the cross-power spectrum of the same galaxies and the matter field of the corresponding DMO simulation.

We compute the model as presented in Eqs. \ref{eq:Pgg} and \ref{eq:Pgm}, leaving the galaxy bias parameters $b_1$, $b_2$, $b_{s^2}$ and $b_{\nabla^2\delta}$, and the amplitude of the shot noise $A_{\rm sn}$ as free parameters. We obtain the spectra for each combination of bias and noise parameters through the \baccoemu{} emulator. In Fig. \ref{fig:coevolution relations} we present the values of $b_1, b_2, b_{s^2}$ and $b_{\nabla^2\delta}$ inferred for the different galaxy samples. Following \cite{ZennaroEtal2022}, we expect higher-order galaxy bias parameters to be correlated with the value of $b_1$. Therefore we compare the values obtained in this work with the coevolution relations from \cite{ZennaroEtal2022}, finding that these empirical fitting functions (with their associated allowed parameter space) encompass all the bias values inferred from the galaxy samples from the \flamingo{} simulations.

\begin{figure*}
    \centering
    \includegraphics[width=\textwidth]{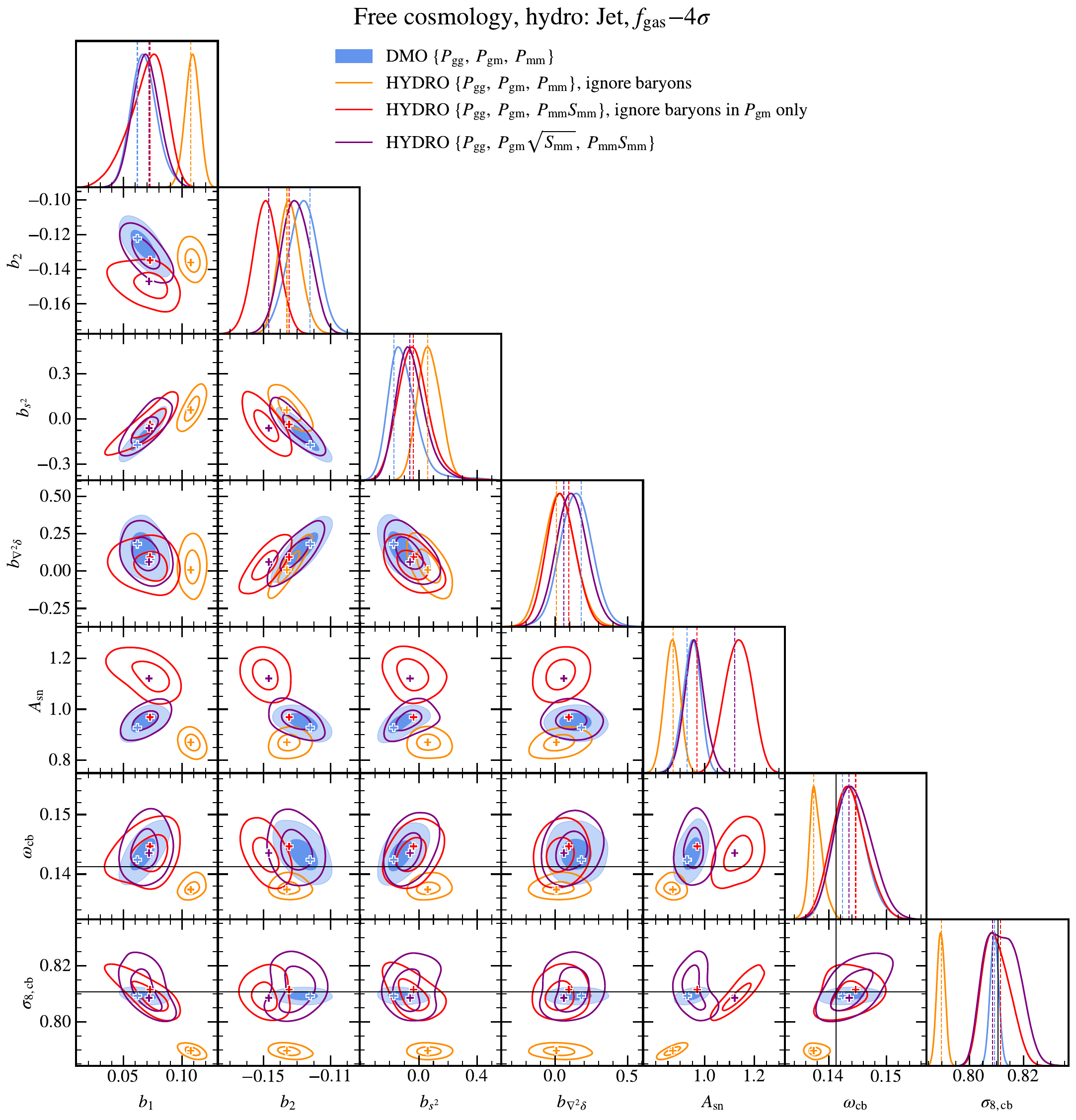}
    \caption{The 1- and 2-$\sigma$ contours of the posterior distribution obtained by fitting the clustering data ($P_{\rm gg}$, $P_{\rm gm}$, and $P_{\rm mm}$) for the "Jet, $f_{\rm gas} - 4\sigma$" simulation at $z=0$, with SM-selected galaxies with number density $n=10^{-2} ~ h^3 ~ \mathrm{Mpc}^{-3}$. Crosses (in the 2D contours) and dotted lines (in the 1D distributions) mark the position of the best-fitting parameters. Here, we jointly fit the galaxy auto and cross-power spectra and the matter auto power spectrum, leaving two cosmological parameters free. Different colours correspond to different cases: \textit{blue} corresponds to the reference values of the bias parameters, where the data come from galaxies (from the hydrodynamical simulation) correlated with matter from the DMO simulation, and the model does not include baryons; in \textit{orange}, galaxies in the data vector are cross correlated with the matter of the hydrodynamical simulation, but the model does not account for baryons, neither in $P_{\rm gm}$ nor in $P_{\rm mm}$; baryons are present in the data also in \textit{red} and are accounted for in the model only in the matter auto-power spectrum, computed as $P_{\rm mm} S_{\rm mm}$; finally, in \textit{purple}, baryons are included in the model both in the galaxy-matter cross-power spectrum, computed as $P_{\rm gm}\sqrt{S_{\rm mm}}$, and in the matter auto-power spectrum, computed as $P_{\rm mm}S_{\rm mm}$. Black lines mark the true values of the cosmological parameters from the simulation. The projection effects visible in some of the contours are discussed in Sec. \ref{sec:projeffs}.}
    \label{fig:freecosmo-pggpgmpmm}
\end{figure*}

\subsection{Effects on bias parameters at fixed cosmology}\label{sec:fixedcosmo}

We now compare the posteriors obtained from fitting our galaxy samples and matter power spectra, comparing cases accounting for or neglecting the effects of baryons on the galaxy-matter power spectrum at fixed cosmology. In this case, we will only show the results for the "Jet, $f_{\rm gas} - 4\sigma$" hydrodynamical simulation at $z=0$, which exhibits the largest suppression at $k=1~\ihMpc$ (see Fig. \ref{fig:flamingo-Sk}). Also, we focus on SM-selected galaxies with number density $n=10^{-2} ~ h^3 ~ \mathrm{Mpc}^{-3}$ in order not to be affected by shot noise.

In Fig. \ref{fig:fixedcosmo}, we present the posteriors obtained when fitting different cases at a fixed cosmology, with and without baryons. Specifically, we consider as a benchmark the case in which our data vector is composed of $\lbrace P_{\rm gg}, P_{\rm gm, dmo} \rbrace$, i.e. the galaxy matter cross-power spectrum is computed using the dark matter field of the DMO simulation, as described in the previous subsection. In this case, we fit our data vector with our reference model for dark matter, in which both the model for $P_{\rm gg}$ and the model for $P_{\rm gm,dmo}$ only depend on matter, with no baryonic effects included. We expect these to be the `true' bias parameters describing this galaxy sample. This reference case corresponds to the blue contours in Fig. \ref{fig:fixedcosmo}.

We then compute the galaxy matter cross-power spectrum $P_{\rm gm}(k)$ with the matter field from the hydrodynamical simulation (orange colour in Fig. \ref{fig:fixedcosmo}). When fitting this new data vector with the same purely DM model, we can still obtain an excellent fit to the data (with $\chi^2$ comparable with the previous case). Still, we find different values for the bias parameters (especially $b_2$ and $A_{\rm sn}$). Note that this implies an inconsistency between the bias parameters that would be preferred by $P_{\rm gg}$ alone (same as the fits obtained for the DMO case) and the bias parameters needed to absorb the effects of baryons in $P_{\rm gm, hydro}$.

If we modify the model by multiplying the theory cross-power spectrum by the baryon suppression computed with the best fitting baryon parameters inferred earlier ($P_{\rm gm,dmo} \sqrt{S_{\rm mm}}$, with $S_{\rm mm}$ computed with the best fitting BCM parameters), we can recover the bias parameters obtained in our fiducial DMO case, finding posteriors in good agreement (green colour). Finally, we have checked that, if we also leave the BCM parameters free (in this case, also fitting the matter power spectrum), then our reference bias parameters are recovered (purple colour).

\begin{figure*}
    \includegraphics[width=\textwidth]{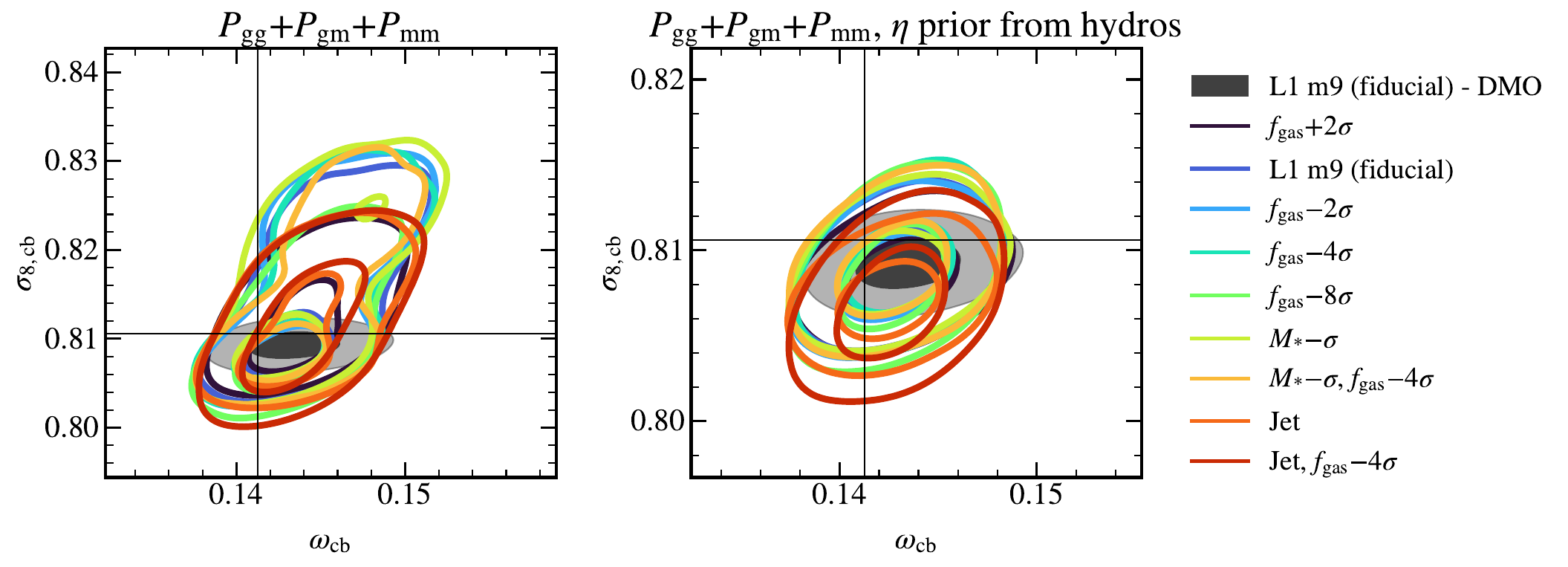}
    \caption{Constraints on the cosmological parameters considered, for all the baryon physics models, with SM-selected galaxies with number density $n=10^{-2} ~ h^3 ~ \mathrm{Mpc}^{-3}$ at $z=0$. The data vector includes the galaxy auto-power spectrum, the cross galaxy-matter power spectrum (both to $k_{\rm max} = 0.7 ~ \ihMpc$) and the matter power spectrum (to $k_{\rm max} = 5 ~ \ihMpc$), with baryonic effects included in both the cross and matter power spectra. In the reference case (dark grey contours), baryonic effects are not present either in the data vector or in the model. We have checked that, by using the DMO simulation with inverted initial phases, the shift of the reference DMO case from the true values of the cosmological parameters can be removed; for the baryonic cases, we therefore focus on shifts between the reference values and each baryonic case. \textit{Left}: all free parameters span the fiducial priors. \textit{Right}: on the baryon model parameter $\eta$ we impose a tight $\log_{10} \eta = -0.32\pm 0.22$ prior, based on 74 hydrodynamical simulations \citep{GrandisEtal2024}. As discussed in Sec. \ref{sec:projeffs}, $\eta$ can become degenerate with $\sigma_8$ when the baryon suppression is weak. Using a phenomenologically motivated prior breaks the degeneracy, and we recover the reference cosmology.}
    \label{fig:logetaprior}
\end{figure*}

\subsection{Effects on bias and cosmological parameters}\label{sec:freecosmo}

From Fig. \ref{fig:fixedcosmo}, we have seen that failing to model the effect of baryons on the galaxy matter cross-power spectrum leads to shifts in the inferred bias parameters. In this section, we investigate whether such a shift, when the cosmological parameters are also left free to vary, can result in incorrect cosmological inferences.

In Fig. \ref{fig:freecosmo-pggpgmpmm}, we repeat the analysis of the previous subsection, but always including the matter power spectrum in our data vector and varying the cosmological parameters $\omega_{\rm cb}$ and $\sigma_{8,\mathrm{cb}}$. Once again, we use as a reference case (blue colour in Fig. \ref{fig:freecosmo-pggpgmpmm}) the one in which both the data vector and the model correspond to the DMO case. When we consider the data vector, including baryon effects, but do not model them (neither for the cross-power spectrum nor the matter power spectrum), we not only find shifts in the inferred bias parameters but also in the cosmological parameters (orange colour in Fig. \ref{fig:freecosmo-pggpgmpmm}). Specifically, $\sigma_8$ in this case is incompatible with the true cosmology of the simulation at the 6-$\sigma$ level. When modelling baryons only at the matter power spectrum level, we can constrain the BCM parameters and find unbiased cosmological parameters, but we once again (similarly to the second case in Sec. \ref{sec:fixedcosmo}) find shifts in the bias parameters (red colour). 

Finally, when modelling baryons both at the matter and cross-power spectrum level, we are able to not only find unbiased cosmological parameters but also recover galaxy bias parameters compatible with our fiducial case (purple colour). In this case, we find important projection effects that affect the shape of our posteriors and broaden the size of our contours. We explore this in the following subsection.

\subsection{Cosmology fits and projection effects}\label{sec:projeffs}
We present now the marginalised posteriors relative to the cosmological parameters in our analysis for all the baryonic models (once again, for simplicity, only considering the SM selected sample with $n=10^{-2} ~ h^3 ~ \mathrm{Mpc}^{-3}$ at $z=0$). As reference, we also show that the constraints on the cosmological parameters obtained for the fiducial model cross correlated with the matter from the DMO simulation, with no baryons neither in the data vector nor in the model (dark grey contours).

These contours are shown in the left panel of Fig. \ref{fig:logetaprior}. We find that in all cases, independently of the baryonic physics considered, including the baryon suppression in the model for the galaxy-matter cross-power spectrum and the matter auto-power spectrum is enough to recover unbiased cosmological results. Specifically, both the $\sigma_{8,\mathrm{cb}}$ and $\omega_{\rm cb}$ values inferred are always compatible within 1$\sigma$ with the reference values obtained in the DMO case. In turn, the reference fiducial DMO case is roughly 1.5$\sigma$ away from the actual simulation values. Using the \flamingo{} DMO simulation with inverted initial phases, we have checked that we can remove this shift, which is, therefore, purely due to cosmic variance. However, the marginalised contours exhibit important non-Gaussianities interpretable as projection effects, including an accentuated bimodality in $\sigma_{8, \mathrm{cb}}$. More specifically, models with weaker baryonic suppressions exhibit a stronger bimodality and models with stronger baryonic suppression have a more Gaussian posterior.

Upon inspection of the full posteriors (see Appendix \ref{appendix:projection}), we concluded that the BCM parameter $\eta$, when left free along with the galaxy bias parameters, exhibits a bimodality that can percolate to $\sigma_{8, \mathrm{cb}}$.
This is not completely unexpected since $\eta$ regulates the AGN feedback distance range, which modulates the suppression of the power spectrum on large scales, without significantly impacting the quantity of gas retained in the haloes, the latter setting the overall amplitude of baryonic effects on the spectrum. $\sigma_{8, \mathrm{cb}}$ varies the overall amplitude of the matter power spectrum but also impacts non-trivially the shape of the baryonic suppression of the spectrum. 

For this reason, we have repeated our analysis assuming a more aggressive prior than the one reported in Tab. \ref{tab:priors}, which was $\log_{10} \eta = \mathcal{U}(-0.7, 0.7)$. Specifically, we follow \cite{GrandisEtal2024}, where a phenomenologically motivated prior $\log_{10}\eta = -0.32 \pm 0.22$ is proposed, based on the values of this parameter inferred from a set of 74 different hydrodynamical simulations \citep{AricoEtal2021b}. As seen from the right panel of Fig. \ref{fig:logetaprior}, with this more stringent prior, we find that the bimodalities in our posteriors disappear. At the same time, we continue to be able to recover the reference values of our cosmological parameters (inferred from the case without baryons) within $1\sigma$, and the actual values (the ones assumed in the simulation) at the 1- or 2-$\sigma$ level (because of cosmic variance).

\section{Conclusions}
In this work, we have assessed the effect of baryons on the galaxy-matter cross-power spectrum using the \flamingo{} suite of cosmological hydrodynamical simulations. We proposed a method to account for this effect in the context of state-of-the-art galaxy clustering and baryon suppression models. Specifically, we focus on modelling galaxy clustering with the hybrid Lagrangian bias expansion models (suited for galaxy-galaxy lensing analyses), and we model the baryon suppression of the matter power spectrum with the so-called baryonification (or Baryon Correction Model, BCM). Our proposed model corrects the galaxy-matter cross-power spectrum computed without accounting for any baryons affecting the matter field, $P_{\rm gm, dmo}(k)$, using the square root of the baryonic suppression inferred from the matter auto power spectrum, $\sqrt{S_{\rm mm}(k)} = [P_{\rm mm, hydro}(k) / P_{\rm mm, dmo}(k)]^{1/2}$, yielding the final galaxy-matter cross-power spectrum in the presence of baryons, $P_{\rm gm, hydro}(k) = P_{\rm gm, dmo}(k) \sqrt{S_{\rm mm}(k)}$. 

While we chose to build our model relying on the emulator suite \baccoemu{} for computing $P_{\rm gm, dmo}(k)$ and $S_{\rm mm}(k)$ as a function of cosmology, galaxy bias, and BCM parameters, we expect any other state-of-the-art galaxy bias and baryon suppression models to perform equally well.

We have performed two types of validations. First, we investigated the performance of the $\sqrt{S_{\rm mm}}$ approximation using measurements of the cross-power spectra from the \flamingo{} hydrodynamical and DMO runs. Second, we showed the performance of the model built with the \baccoemu{} emulator when fitting simulated data from the \flamingo{} simulations.

To this end, we selected various halo and galaxy samples in nine different \flamingo{} baryonic models. In each simulation, we selected galaxies based on their stellar mass or star formation rate. We consider three different number densities, effectively changing the typical halo masses contributing most to each sample. To understand our results, we also considered directly the halo populations of the different \flamingo{} simulations, split into mass bins. While our main analysis focused on redshift $z=0$ (where baryonic effects are largest), we have repeated our analysis at redshift $z=1$ -- which is expected to dominate in upcoming lensing surveys. 

We list here our main conclusions.
\begin{itemize}
    \item We measure the baryonic suppressions (at redshifts $z=0$ and $1$) from the matter power spectrum $S_{\rm mm}(k)$ using nine baryonic models from the \flamingo{} simulations. We show we can fit them to subpercent level using the BCM \baccoemu{} emulator down to $k=5 ~ \ihMpc$ (Fig \ref{fig:flamingo-Sk}).
    \item For haloes of different masses, we show that the baryonic suppression inferred from the matter power spectrum is not always a good description of the suppression of the halo-matter cross-power spectrum (Fig. \ref{fig:sqrtSk-all}). Specifically, low-mass haloes exhibit stronger suppressions of the cross-power spectrum than inferred from the matter alone, while the opposite is true for high-mass haloes. Haloes of mass between $10^{13.5}$ and $10^{14} ~h^{-1} ~ \mathrm{M}_\odot$ exhibit suppressions similar to the one inferred from the matter auto-power spectrum -- a not-so-surprising result since haloes of this mass are the ones principally contributing to the matter power spectrum on the scales of interest here \citep{VanDaalenSchaye2015}. 
    \item Focusing on galaxy samples and scales relevant for hybrid perturbative galaxy bias models ($k < 0.7 ~ \ihMpc$), we find that correcting the galaxy-matter cross-power spectrum with the suppression inferred from the matter auto power spectrum ($\sqrt{S_{\rm mm}(k)}$) is accurate at the percent level, independently from galaxy selection criteria, number density, redshift and the specifics of the baryonic physics considered.
    \item We analyse galaxy samples from the \flamingo{} simulations at redshift $z=0$ (where the baryonic effects are largest) and found that accounting for baryonic effects on $P_{\rm mm}$, but not on $P_{\rm gm}$, does not affect the quality of the fit and delivers cosmological constraints compatible within $1\sigma$ from the reference values (from the non-baryonic case), and within 1-2$\sigma$ from the true values (because of cosmic variance), but leads to a shift in the inferred galaxy bias parameters -- which become incompatible with the ones obtained in a DMO context (by up to 6$\sigma$). Including baryonic effects in the modelling of $P_{\rm gm}$ (by correcting the $P_{\rm gm, dmo}$ predicted for dark matter with an extra $\sqrt{S_{\rm mm}(k)}$ term) allows us to recover the same galaxy bias parameters obtained in the DMO context, as well as cosmological parameters within $1\sigma$ of their reference values. We also notice that ignoring baryonic effects on $P_{\rm gm}$ and $P_{\rm mm}$ leads to biased posteriors for the galaxy bias parameters and the cosmological parameters (once again, by up to $6\sigma$).
    \item Finally, we find that when fitting both the galaxy bias and the BCM parameters, non-trivial degeneracies can arise; these do not bias the inferred parameters but can cause the posteriors to exhibit strong non-Gaussianities. While the posterior for this model could, in principle, be non-Gaussian, we find that physically informed priors on the BCM parameters can break these degeneracies and, in particular, a simple, phenomenologically motivated prior on $\eta$ yields Gaussian posteriors compatible within $1\sigma$ with the reference values, and within 1-2$\sigma$ with the true values (because of cosmic variance).
\end{itemize}
We expect this work to aid in designing analysis pipelines for upcoming stage-IV galaxy surveys and reanalyses of current data sets, including scales usually discarded to avoid baryonic contamination. We note that we focused here on the galaxy-matter cross-power spectrum because the effect of baryons on the galaxy auto power spectrum is expected to be subdominant and entirely captured by the galaxy bias parametrisation. However, we leave investigating the physics of how baryons affect the central and satellite subhalo distribution for a future extension of this work. 

\begin{acknowledgements}
The authors would like to acknowledge David Alonso, Elisa Chisari, and Jaime Salcido for their useful feedback and discussion.
MZ is supported by STFC. CGG is supported by the Beecroft Trust.  SC acknowledges the support of the `Juan de la Cierva Incorporac\'ion' fellowship (IJC2020-045705-I). LO acknowledges the support of ”la Caixa” Foundation (ID 100010434) for the fellowship with code LCF/BQ/DR21/11880028. REA acknowledges support from project PID2021-128338NB-I00 from the Spanish Ministry of Science and support from the European Research Executive Agency HORIZON-MSCA-2021-SE-01 Research and Innovation programme under the Marie Skłodowska-Curie grant agreement number 101086388 (LACEGAL). 
\end{acknowledgements}

\bibliographystyle{aa} 
\bibliography{Bibliography_all} 

\begin{appendix}
\section{Halo and galaxy selection convergence}\label{appendix:res}
In this appendix, we test that our selection criteria for galaxies (and the haloes they populate) are robust. We do so by comparing it against a higher-resolution simulation. The \flamingo{} suite includes realisations with different resolutions for the fiducial cosmology and baryonic model. Therefore we will focus on the "L1 m9" model from Tab. \ref{tab:baryon-names}, considering the same resolution used in this work (dubbed \textit{standard res}), with box size $L=681~\hMpc$ and $N_{\rm dm} = N_{\rm gas} = 1800^3$ and $N_\nu=1000^3$, a larger realisation with same resolution ($L=1906.8 ~\hMpc$, $N_{\rm dm} = N_{\rm gas} = 5040^3$ and $N_\nu=2800^3$, dubbed \textit{large box}), and a high resolution realization with $L=681~\hMpc$ and $N_{\rm dm} = N_{\rm gas} = 3600^3$ and $N_\nu=2000^3$, dubbed \textit{high res}.

\begin{figure*}
    \includegraphics[width=\textwidth]{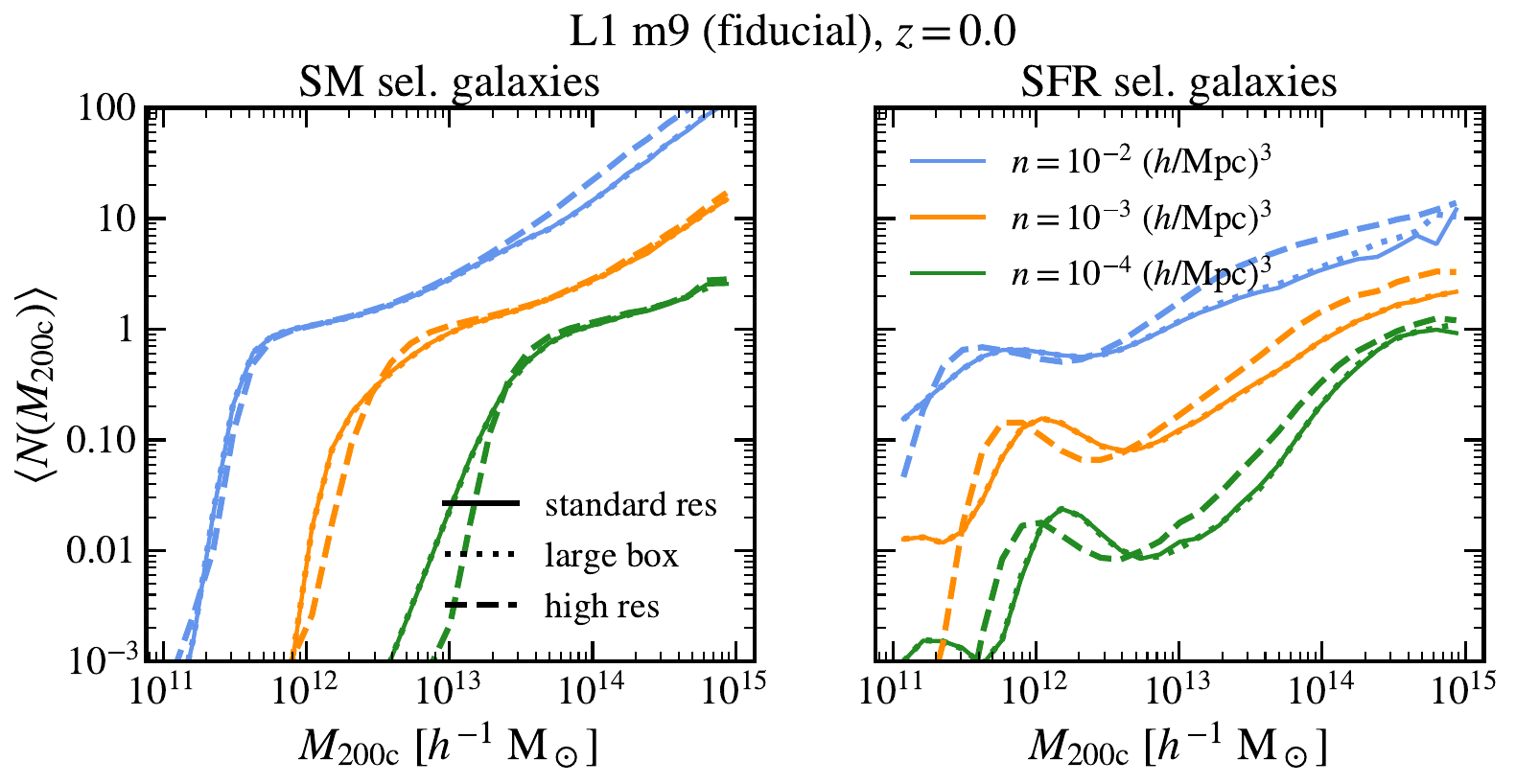}
    \caption{Same as Fig. \ref{fig:gal-hods}, but for the fiducial baryonic model, considering the standard resolution simulation, a larger volume simulation with the same resolution as the standard one, and a higher resolution simulation. For the standard resolution, SM-selected samples with higher number densities lack satellite galaxies with respect to the higher resolution case; also, for SFR-selected galaxies, the standard resolution simulation is significantly lacking satellite galaxies for the high number density samples, and the peak corresponding to the transition from central to satellite galaxies in the HOD is not correctly captured, being systematically shifted towards larger halo masses.}
    \label{fig:hod_hr}
\end{figure*}

In Fig. \ref{fig:hod_hr} we compare the HOD of galaxy samples selected according to the criteria used in this work (either stellar mass or star formation rate, for number densities $n_{\rm g} = \{10^{-2}, 10^{-3}, 10^{-4}\}~h^3 ~ \mathrm{Mpc}^{-3}$) from the "standard res", "large box", and "high res" simulations at $z=0$. The HODs obtained using the two simulations with the same mass resolution but different volumes are virtually identical, indicating that our selection criteria are robust against changes in volume, and the volume considered in our main analysis is sufficient. On the other hand, the "high res" simulation exhibits larger satellite fractions, especially for the denser samples, for both SM and SFR-selected galaxies. Moreover, it exhibits a steeper HOD for central galaxies at small halo masses, especially for the sparsest SM-selected galaxy sample. For SFR-selected galaxies, the HODs from the "high res" simulation also show a significant shift of the characteristic central-to-satellite bump towards smaller halo masses. All these differences point towards the standard resolution simulations not being able to resolve small enough haloes, hosting either low-halo mass centrals or satellites.

This is a very important limitation of our analysis, especially for SFR-selected galaxies, and sparse SM-selected samples. These galaxy samples will necessarily be, to a certain degree, not sufficiently realistic. Nonetheless, our goal in selecting galaxies according to different criteria is \textit{not} to reproduce realistic samples, but to test our model with galaxy samples with very different biasing properties with respect to the underlying dark matter, and populating haloes of very different masses. The samples obtained from the "standard res" simulations, albeit not entirely realistic, do fulfil this purpose and can, therefore, be used for our analysis.

\begin{figure*}
    \includegraphics[width=\textwidth]{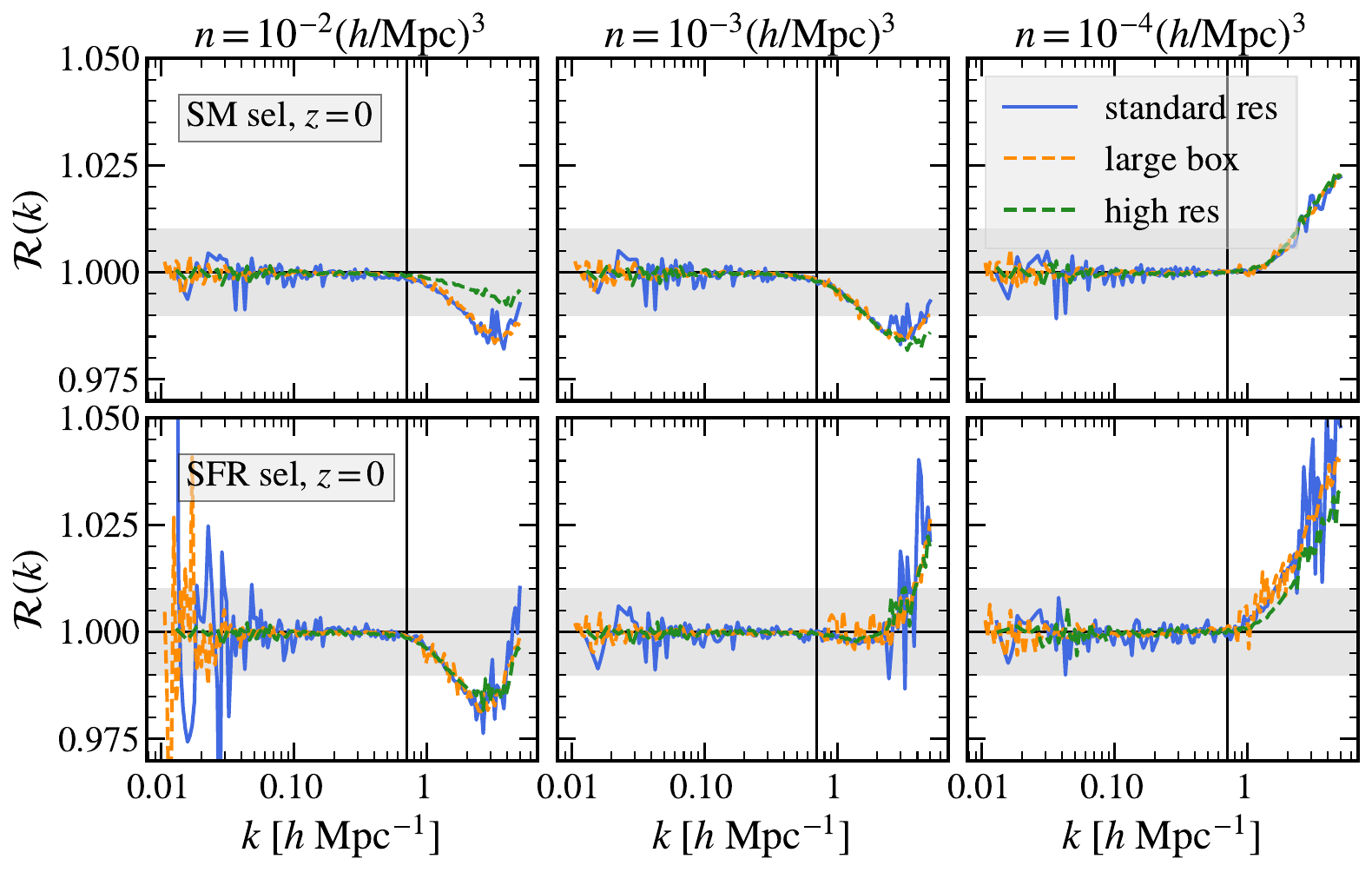}
    \caption{Same as Fig. \ref{fig:sqrtSk-all}, but for the fiducial baryonic model, considering the standard resolution simulation, a larger volume simulation with the same resolution as the standard one, and a higher resolution simulation at $z=0$. For the three configurations considered, the $\sqrt{S_{\rm mm}(k)}$ approximation is 1\% accurate on all relevant scales, irrespective of the galaxy selection criterion of number density.}
    \label{fig:RR_HR}
\end{figure*}

To further prove this point, in Fig. \ref{fig:RR_HR} we repeat our analysis presented in Fig \ref{fig:sqrtSk-all} (at $z=0$), but comparing the "standard res", "large box" and "high res" simulations. We use the same metric $\mathcal{R}$ to assess the accuracy of the $\sqrt{S_{\rm mm}(k)}$ approximation. We find that, on the scales of interest ($k < 0.7 ~ \ihMpc$), the $\sqrt{S_{\rm mm}(k)}$ approximation does not depend on the simulation resolution, always exhibiting sub-percent accuracy. On smaller scales, we find differences when using the "high res" simulations, with the $\sqrt{S_{\rm mm}(k)}$ approximation generally working even better than for the "standard res" case. For this reason, we are confident that the results of our main analysis are not affected by the resolution of the simulations used.

\section{BCM best fitting parameters}
We report in Tab. \ref{tab:Smmz0-bf} and \ref{tab:Smmz1-bf} the best fitting parameters of the BCM, inferred by fitting the measured $S_{\rm mm}(k)$, respectively at $z=0$ and $z=1$, for the different hydrodynamical simulations considered. Since our covariance matrices are just an approximation (see Sec. \ref{sec:likelihood}), we only quote the best-fitting values and not credibility intervals.

\begin{table*}
    \centering
        \begin{tabular}{|l|r|r|r|r|r|r|r|}
        \hline
        baryon model & $\log_{10}(M_{\rm c})$ & $\log_{10}(\eta)$ & $\log_{10}(\beta)$ & $\log_{10}(M_{1,z_0,\mathrm{cen}})$ & $\log_{10}(\theta_{\rm inn})$ & $\log_{10}(M_{\rm inn})$ & $\log_{10}(\theta_{\rm out})$ \\
        \hline
        $f_{\rm gas}$+2$\sigma$ & 12.8797 & -0.0891 & 0.0431 & 9.9021 & -0.5317 & 13.0030 & 0.0065 \\
        L1 m9 (fiducial) & 13.3294 & -0.3941 & -0.0582 & 9.8245 & -0.5680 & 12.6675 & 0.0057 \\
        $f_{\rm gas}$$-$2$\sigma$ & 13.1060 & -0.5376 & -0.4560 & 9.9945 & -0.5340 & 13.0193 & 0.0261 \\
        $f_{\rm gas}$$-$4$\sigma$ & 13.4830 & -0.5229 & -0.5056 & 10.5330 & -0.5794 & 12.6545 & 0.0066 \\
        $f_{\rm gas}$$-$8$\sigma$ & 14.2525 & -0.2922 & 0.4050 & 11.8526 & -0.6878 & 11.5687 & 0.4539 \\
        $M_{*}$$-$$\sigma$ & 13.0003 & -0.5296 & -0.4093 & 10.1514 & -0.5856 & 13.3151 & 0.0007 \\
        $M_{*}$$-$$\sigma$, $f_{\rm gas}$$-$4$\sigma$ & 13.5454 & -0.4906 & -0.4927 & 10.4512 & -0.5364 & 12.9636 & 0.0072 \\
        Jet & 13.3950 & -0.0806 & 0.3008 & 10.5697 & -0.5414 & 11.8883 & 0.0256 \\
        Jet, $f_{\rm gas}$$-$4$\sigma$ & 14.0573 & -0.1892 & 0.0596 & 10.3642 & -0.8323 & 10.0646 & 0.0994 \\
        \hline
        \end{tabular}
    \caption{Best fitting parameters of the BCM model for the different baryonic scenarios at $z=0$. Fits are performed for $k \leq 5 ~ \ihMpc$. Note that, for typesetting reasons, we abused the notation and $\log_{10}(M_{\rm c})$, $\log_{10}(M_{1,z_0,\mathrm{cen}})$, and $\log_{10}(M_{\rm inn})$ should be $\log_{10}[M_{\rm c} / (h^{-1} ~ \mathrm{M}_{\odot})]$, $\log_{10}[M_{1,z_0,\mathrm{cen}} / (h^{-1} ~ \mathrm{M}_{\odot})]$, and $\log_{10}[M_{\rm inn} / (h^{-1} ~ \mathrm{M}_{\odot})]$.}
    \label{tab:Smmz0-bf}
\end{table*}

\begin{table*}
    \centering
        \begin{tabular}{|l|r|r|r|r|r|r|r|}
        \hline
        baryon model & $\log_{10}(M_{\rm c})$ & $\log_{10}(\eta)$ & $\log_{10}(\beta)$ & $\log_{10}(M_{1,z_0,\mathrm{cen}})$ & $\log_{10}(\theta_{\rm inn})$ & $\log_{10}(M_{\rm inn})$ & $\log_{10}(\theta_{\rm out})$ \\
        \hline
        $f_{\rm gas}$+2$\sigma$ & 13.6519 & -0.6472 & 0.6354 & 10.5423 & -1.6187 & 9.5013 & 0.0700 \\
        L1 m9 (fiducial) & 14.1718 & -0.6845 & -0.0098 & 11.9372 & -1.0348 & 11.2650 & 0.4395 \\
        $f_{\rm gas}$$-$2$\sigma$ & 14.9928 & -0.6852 & -0.1889 & 12.0781 & -1.6774 & 10.2274 & 0.1826 \\
        $f_{\rm gas}$$-$4$\sigma$ & 14.9921 & -0.5916 & -0.1112 & 12.1247 & -1.5308 & 9.7409 & 0.4068 \\
        $f_{\rm gas}$$-$8$\sigma$ & 14.9758 & -0.4872 & 0.5980 & 12.3158 & -0.9959 & 9.3944 & 0.0018 \\
        $M_{*}$$-$$\sigma$ & 14.8618 & -0.6797 & -0.6690 & 9.6316 & -1.8848 & 11.0208 & 0.3665 \\
        $M_{*}$$-$$\sigma$, $f_{\rm gas}$$-$4$\sigma$ & 14.9668 & -0.5683 & -0.0734 & 11.9403 & -1.5656 & 9.5252 & 0.0587 \\
        Jet & 13.8292 & -0.5783 & 0.3203 & 11.9680 & -1.3738 & 11.0009 & 0.1381 \\
        Jet, $f_{\rm gas}$$-$4$\sigma$ & 14.9772 & -0.2948 & 0.3497 & 12.6713 & -1.1232 & 10.2000 & 0.2877 \\
        \hline
        \end{tabular}
    \caption{As Tab. \ref{tab:Smmz0-bf}, but for $z=1$.}
    \label{tab:Smmz1-bf}
\end{table*}

\section{Higher redshift}
We repeat the cosmology fixed analysis of Fig. \ref{fig:fixedcosmo} at $z=1$. We see from Fig. \ref{fig:flamingo-Sk} that baryon effects in these simulations induce smaller suppression (on the matter power spectrum) as we move to higher redshift. We therefore expect the $\sqrt{S_{\rm mm}(k)}$ correction to hold also for higher redshifts. We consider here $z=1$, which corresponds roughly to the redshifts at which the redshift distributions of upcoming surveys (such as Euclid or Rubin-LSST) are expected to peak \citep{EuclidValidation2020, IlbertEtal2021,EuclidPresentation2024,NicolaEtal2024}.

Once again, we focus on the "Jet, $f_{\rm gas} - 4\sigma$"  hydrodynamical simulation, where we select galaxies according to their stellar mass and retain a number density cut catalogue with $\bar{n} = 10^{-2} ~ h^3 ~ \mathrm{Mpc}^{-3}$

\begin{figure*}
    \centering
    \includegraphics[width=\textwidth]{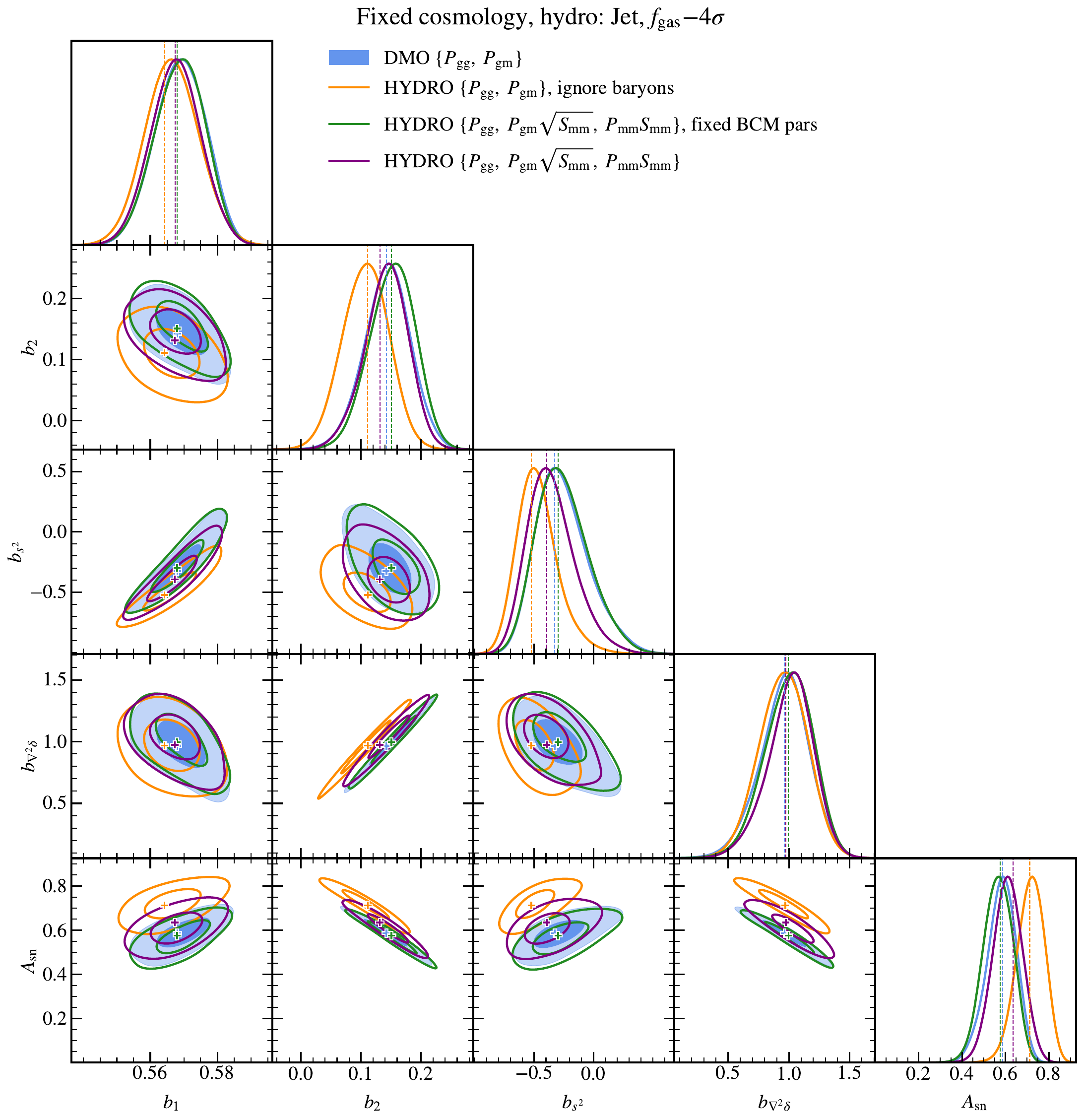}
    \caption{As Fig. \ref{fig:fixedcosmo}, but for $z=1$.}
    \label{fig:fixedcosmo-z1}
\end{figure*}

Our results are presented in Fig. \ref{fig:fixedcosmo-z1}. We find that the $S_{\rm mm}(k)$ correction can still deliver unbiased results. The trends are exactly the same ones as in the $z=0$ case: when ignoring the effect of baryons in the model for the galaxy-matter cross-power spectrum, the galaxy bias parameters assume values different from their reference ones, allowing us to still find an excellent fit to the data, but yielding inconsistent bias parameters between the auto and cross-power spectrum. Not surprisingly, this inconsistency is milder at this higher redshift and below the 2-$\sigma$ significance in all cases. When including the $\sqrt{S_{\rm mm}}$ term in the modelling of the galaxy matter cross-power spectrum, we obtain values of the galaxy bias parameters compatible with the reference ones, both if we fix the BCM parameters to their best fitting values shown in Tab. \ref{tab:Smmz1-bf}, and if we leave the BCM parameters free.

\section{Degeneracy between galaxy bias and BCM parameters}\label{appendix:projection}
In Fig. \ref{fig:logetaprior}, we showed how leaving all galaxy bias and BCM parameters free can introduce a bimodality in the BCM parameter $\eta$, that is reflected in the cosmological parameter $\sigma_{8,\mathrm{cb}}$. 

\begin{figure*}
    \centering
    \includegraphics[width=\textwidth]{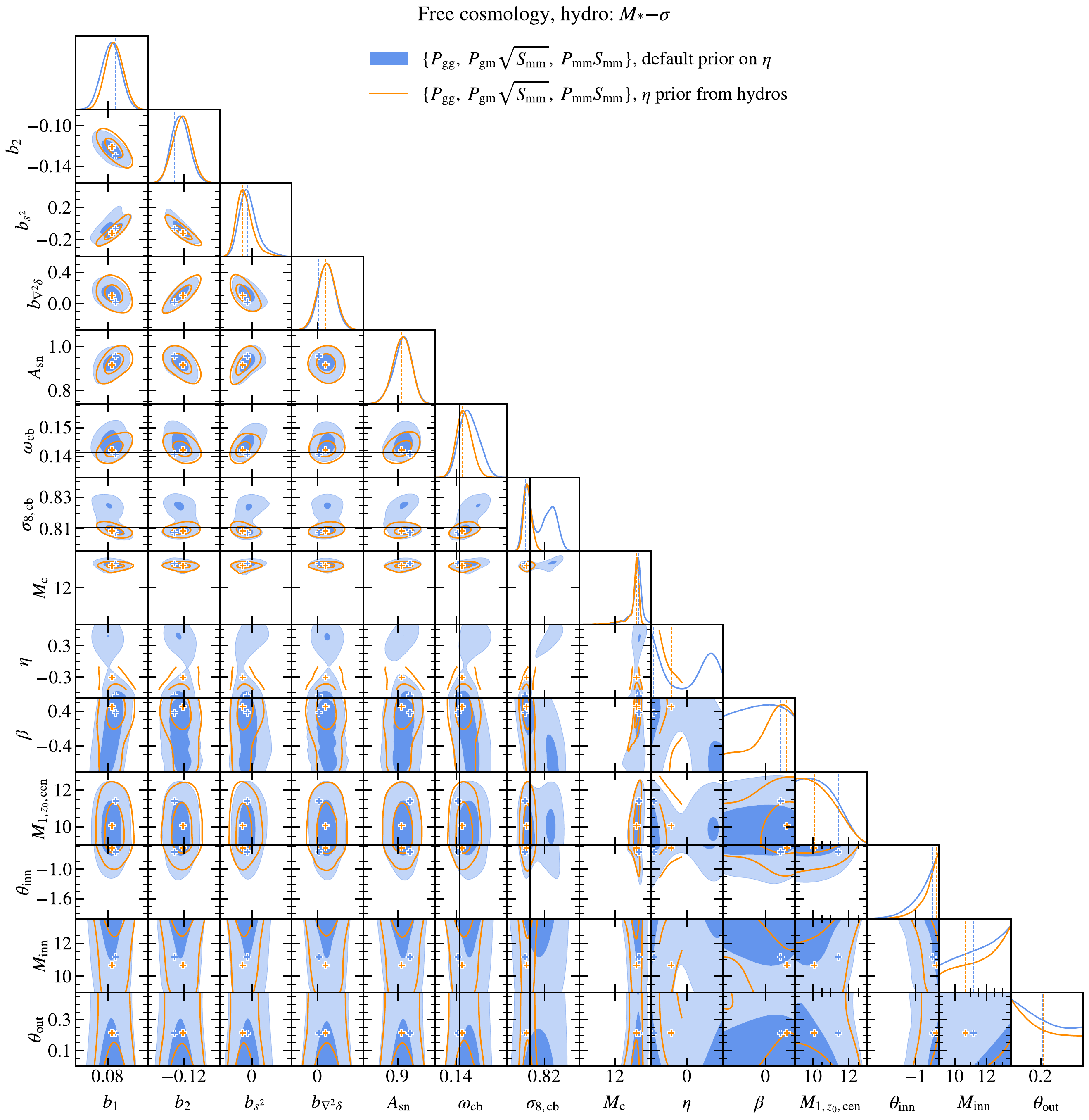}
    \caption{The 1- and 2-$\sigma$ 2D contours of the posterior distribution obtained fitting the clustering data for the "$M_{*} - \sigma$" simulation at $z=0$, with SM-selected galaxies with number density $n=10^{-2} ~ h^3 ~ \mathrm{Mpc}^{-3}$. Blue contours correspond to using the priors of Tab. \ref{tab:priors}, and orange contours correspond to setting a tighter prior on the BCM parameter $\eta$. Note that for space reasons, all the BCM parameters in the plot have misleading labels since the free parameters are not $M_{\rm c}$, etc \ldots, but their logarithms ($\log_{10}[M_{\rm c} / (h^{-1} ~ \mathrm{M}_\odot)]$, etc\ldots).}
    \label{fig:degeneracies}
\end{figure*}

We make this explicit in Fig. \ref{fig:degeneracies}, where we show the posteriors obtained fitting the full data vector from the "$M_{*} - \sigma$" hydrodynamical simulation (SM-selected galaxies at $z=0$ with $\bar{n}=10^{-2}~h^3~\mathrm{Mpc}^{-3}$). We chose this baryonic model because its posterior exhibits the largest bimodality in Fig. \ref{fig:logetaprior}. 

We consider the same case as before, including baryonic effects in the galaxy-matter cross-power spectrum and the matter auto-power spectrum, and using the same priors as in Tab. \ref{tab:priors}. We see that indeed $\log_{10}\eta$ shows non-trivial degeneracies with the galaxy bias parameters, and we find a large bimodality with positive and negative values of $\log_{10} \eta$ being almost equally likely.

Once we impose the additional prior on $\eta$ from \cite{GrandisEtal2024}, namely $\log_{10}\eta = -0.32 \pm 0.22$, we remove the bimodality in $\eta$ (and consequently $\sigma_{8, \mathrm{cb}}$), but recover virtually the same posteriors for the other parameters.

\end{appendix}

\end{document}